\documentclass[sigconf]{acmart}

\usepackage{hyperref}
\usepackage{fancyhdr}

\usepackage{algorithmic}
\usepackage{graphicx}
\usepackage{textcomp}
\usepackage{xcolor}

\usepackage{booktabs}

\usepackage{tabularx}  
\usepackage{booktabs} 
\usepackage{makecell}

\usepackage{bbding}
\usepackage{multirow}
\usepackage{caption}
\usepackage{subcaption} 
\usepackage{siunitx}
\usepackage{csquotes}
\usepackage{enumitem}
\usepackage{comment}
\usepackage{color}
\usepackage{colortbl}
\usepackage{soul}

\usepackage{lipsum}

\newcommand{\oursys}{{\textsc{LUT Tensor Core}}\xspace}

\usepackage[linesnumbered,ruled,vlined]{algorithm2e}
\usepackage{todonotes}

\AtBeginDocument{%
  }

\copyrightyear{2025} 
\acmYear{2025} 
\setcopyright{cc}
\setcctype{by-nc-nd}
\acmConference[ISCA '25]{Proceedings of the 52nd Annual International
 Symposium on Computer Architecture}{June 21--25, 2025}{Tokyo, Japan}
\acmBooktitle{Proceedings of the 52nd Annual International Symposium on
 Computer Architecture (ISCA '25), June 21--25, 2025, Tokyo,
 Japan}
\acmDOI{10.1145/3695053.3731057}
\acmISBN{979-8-4007-1261-6/2025/06}

\begin{document}

\title{\oursys{}: A Software-Hardware Co-Design for LUT-Based Low-Bit LLM Inference}

% \textsuperscript{\dag}
\author{Zhiwen Mo}
\authornote{Work is done during internship at Microsoft Research.}
\affiliation{%
  \institution{Imperial College London and Microsoft Research}
  \country{London, UK}
}

\author{Lei Wang}
\authornotemark[1]
\affiliation{%
  \institution{Peking University and\\ Microsoft Research}
  \country{Beijing, China}
}

\author{Jianyu Wei}
\authornotemark[1]
\affiliation{%
  \institution{University of Science and Technology of China and Microsoft Research}
  \country{Beijing, China}
}

\author{Zhichen Zeng}
\authornotemark[1]
\affiliation{%
  \institution{University of Washington and Microsoft Research}
  \country{Seattle, USA}
}

\author{Shijie Cao}
\authornote{Corresponding Author}
\affiliation{%
  \institution{Microsoft Research}
  \country{Beijing, China}
}

\author{Lingxiao Ma}
\affiliation{%
  \institution{Microsoft Research}
  \country{Beijing, China}
}

\author{Naifeng Jing}
\affiliation{%
  \institution{Shanghai Jiao Tong University}
  \country{Shanghai, China}
}

\author{Ting Cao}
\affiliation{%
  \institution{Microsoft Research}
  \country{Beijing, China}
}

\author{Jilong Xue}
\affiliation{%
  \institution{Microsoft Research}
  \country{Beijing, China}
}

\author{Fan Yang}
\affiliation{%
  \institution{Microsoft Research}
  \country{Beijing, China}
}

\author{Mao Yang}
\affiliation{%
  \institution{Microsoft Research}
  \country{Beijing, China}
}

\renewcommand{\shortauthors}{Mo et al.}

\begin{abstract}

Large Language Model (LLM) inference becomes resource-intensive, prompting a shift toward low-bit model weights to reduce the memory footprint and improve efficiency.
Such low-bit LLMs necessitate the mixed-precision matrix multiplication (mpGEMM), an important yet underexplored operation involving the multiplication of lower-precision weights with higher-precision activations. 
Off-the-shelf hardware does not support this operation natively, leading to indirect, thus inefficient, dequantization-based implementations.

In this paper, we study the lookup table (LUT)-based approach for mpGEMM and find that a conventional LUT implementation 
fails to achieve the promised gains.
To  unlock the full potential of LUT-based mpGEMM, we propose \oursys{}, a software-hardware co-design for low-bit LLM inference. 
\oursys{} differentiates itself from conventional LUT designs through: 
1) software-based optimizations to minimize table precompute overhead and weight reinterpretation to reduce table storage;
2) a LUT-based Tensor Core hardware design with an elongated tiling shape to maximize table reuse and a bit-serial design to support diverse precision combinations in mpGEMM; 
3) a new instruction set and compilation optimizations for LUT-based mpGEMM.
\oursys{} significantly outperforms existing pure software LUT implementations and achieves a 1.44$\times$ improvement in compute density and energy efficiency compared to previous state-of-the-art LUT-based accelerators.

\end{abstract}




\maketitle

\section{Introduction}
\label{sec:introduction}
The advent of Large Language Models (LLMs) offers transformative opportunities across various AI applications~\cite{brown2020language,achiam2023gpt,kaplan2020scaling,touvron2023llama}. 
However, the deployment of LLMs requires substantial hardware resources~\cite{hoffmann2022training,patterson2022carbon,patel2023splitwise}.
To reduce inference costs, low-bit LLMs have emerged as promising approaches~\cite{frantar2022gptq,kim2023squeezellm,liu2024qllm,dettmers2022gpt3.int8}. 
Among different solutions, weight quantization, i.e., quantizing LLMs with low-precision weights and high-precision activations, has become particularly attractive as it reduces memory and computation costs while maintaining model accuracy~\cite{lin2023awq,yao2022zeroquant,zhang2023integer}. 
While 4-bit weight quantization has become pervasive~\cite{dettmers2023case,team2023gemini,kumar2024scaling}, 
both academia and industry are actively 
exploring advancements
toward 2-bit and even 1-bit to further improve efficiency~\cite{chee2024quip, wang2023bitnet,du2024bitdistiller, liu2025paretoq,nair2025matryoshka,ma2024era,kaushal2024spectra}.

Weight quantization shifts the key computation pattern of LLM inference from conventional General Matrix Multiplication (GEMM) to \textbf{mixed-precision GEMM (mpGEMM)}, where one input matrix is in lower precision (e.g., INT4/2/1 weights) and the other remains in higher precision (e.g., FP16/8, INT8 activations).
Currently, off-the-shelf hardware does not support mixed-precision operations \textit{natively}.
Consequently, most low-bit LLM inference systems have to utilize 
\textit{dequantization-based} approaches for mpGEMM~\cite{lin2023awq,trt-llm,llamacpp,wang2024ladder}. Dequantization upscales low-bit representations to match the hardware-supported GEMM.
Such extra operations can become a performance bottleneck in large batch scenarios.

Lookup table (LUT) is another popular approach for low-bit computation and is well-suited for mpGEMM~\cite{jeon2020biqgemm,maleki2023look,park2023lutgemm,wei2024t,unpu}.
By precomputing multiplication results between low-precision weights and high-precision activations, LUT-based methods eliminate the need for dequantization and replace complex operations with simple table lookups.
In practice, LUTs are implemented on a per-tile basis. For each small tile of mpGEMM, a lookup table is precomputed specifically for the activations within a tile and reused across weight matrix columns, significantly reducing storage overhead while maintaining efficiency.

Despite its potential, LUT-based mpGEMM still experiences notable performance gaps and challenges in both software and hardware implementations.
On the software side, LUT kernels face limited instruction support and inefficient memory access, which lead to suboptimal performance compared to dequantization-based kernels on GPUs, as shown in Figure~\ref{fig:motivation}. 
On the hardware side, conventional LUT designs 
lack optimization 
for mpGEMM and often fall short of expected
performance improvements. This is due to key challenges such as high table precomputation and storage overhead, limited support for diverse bit-width combinations, inefficiencies from suboptimal tiling shapes, and the lack of dedicated instruction sets and compilation support; see \S\ref{ssec:challenges} for details.

\oursys{} addresses these challenges through a holistic software and hardware co-design.
By optimizing hardware-unfriendly tasks, such as table precomputation and storage management, in a software-based approach, \oursys{} reduces the workload on hardware, simplifying its design and improving its compactness and efficiency.
To be specific:

\noindent
\textbf{Software optimization (\S~\ref{ssec:design_software})}.
To amortize the overhead of precomputing lookup tables, we observed that conventional designs precompute redundantly across multiple units. \oursys{} splits the precomputation into an independent operator, thus avoiding redundancies, and fuses it with the previous operator to further reduce memory accesses.
To reduce storage overhead, \oursys{} exposes and exploits the inherent symmetry of a lookup table for mpGEMM by reinterpreting $\{0, 1\}$ as $\{-1, 1\}$, reducing the table size by half. 
\oursys{} also reduces the table width and supports various activation bit widths by applying table quantization techniques. 

\noindent
\textbf{Hardware customization (\S~\ref{ssec:design_hardware})}.
\oursys{} customizes the LUT-based Tensor Core design.  
The aforementioned software optimizations have simplified the hardware design by offloading the circuitry tasks to software, reducing the need for broadcasting and multiplexers by half. Meanwhile, \oursys{} incorporates a flexible bit-serial-like circuit to accommodate various combinations of mixed precision operations.
Moreover, \oursys{} conducts a design space exploration (DSE) for the shape of LUT-based Tensor Core and identifies an elongated tiling shape that enables more efficient table reuse.

\noindent
\textbf{New instruction and compilation support (\S~\ref{ssec:design_compilation})}.
\oursys{} extends the traditional Matrix Multiply-Accumulate (MMA) instruction set to the LUT-based Matrix Multiply-Accumulate (LMMA) instruction set, which includes essential metadata specifying the operand types and shapes. 
With the extension, \oursys{} leverages the shape information provided in LMMA to recompile LLM workloads using tile-based deep learning compilers~\cite{chen2018tvm,zhu2022roller,shi2023welder}, producing efficient
kernels for the new hardware.

Our LUT-based Tensor Core exhibits a power and area reduction of 4$\times$ to 6$\times$ compared to the conventional Tensor Core.
To validate the performance enhancement of mpGEMM, we integrate  
our LUT-based Tensor Core design and instructions 
into Accel-Sim~\cite{khairy2020accel}, a GPU hardware simulator. The results show that our LUT-based Tensor Core occupies only 16\% of the area of a conventional Tensor Core while achieving even higher mpGEMM performance.
Compared to state-of-the-art (SOTA) LUT software implementations \cite{park2023lutgemm}, \oursys{} achieves up to a 1.42$\times$ speedup in general matrix vector multiplications (GEMV) and a 72.2$\times$ speedup in GEMM. Compared to SOTA LUT accelerators~\cite{unpu}, \oursys{} achieves 1.44$\times$ higher compute density and energy efficiency, enabled by the software-hardware co-design.
Our code is available at \url{https://github.com/microsoft/T-MAC/tree/LUTTensorCore_ISCA25}.

Our contributions can be summarized as follows:
\vspace{-1mm}
\begin{itemize}[leftmargin=*]
    \item 
    We propose \oursys{}, a software-hardware co-design for LUT-based mpGEMM to boost the inference efficiency of low-bit LLMs.

    \item Experiments show that the proposed LUT-based Tensor Core achieves  4$\times$ to 6$\times$ power, performance, and area (PPA) gains. \oursys{} exhibits  inference speedups of 2.06$\times$ to 5.51$\times$ for low-bit LLMs like BitNet and quantized LLAMA models, with comparable area and accuracy.

    \item Beyond efficiency, our design can accommodate a wide range of weight (e.g., INT4/2/1) and activation precisions (e.g., FP16/8, INT8). Moreover, \oursys{} can be integrated into existing inference hardware and software stacks with the extended LMMA instructions and compilation optimizations.
   
\end{itemize}
\section{Background and Motivation} \label{sec:background}
\label{sec:background_and_motivation}

\subsection{LLM Inference and Low-Bit Quantization} \label{ssec:bg}

\begin{figure}[t]
    \centering
    \includegraphics[width=0.83\linewidth]{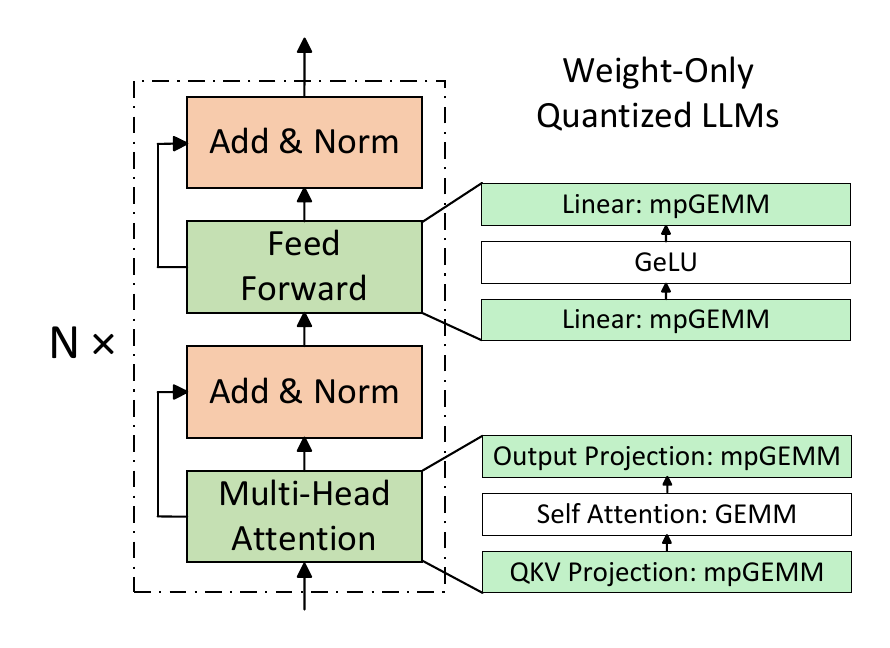}
    \vspace{-3mm}
    \caption{Decoder-only transformer blocks in LLMs. The primary computations are GEMM operations (or mpGEMM operations with weight quantization).}
    \vspace{-3mm}
    \label{fig:llm}
\end{figure}

Recently, LLMs mainly rely on the decoder-only transformer architecture~\cite{vaswani2017attention}, as shown in Figure~\ref{fig:llm}.
Specifically, LLMs are built with sequential transformer layers, where each transformer layer contains a multi-head attention block followed by a feed-forward block. In both blocks, the primary computations are GEMM, or mpGEMM operations with weight quantization.
The scaling up of LLMs requires substantial hardware resources~\cite{kaplan2020scaling,hoffmann2022training}. For example, LLAMA-2-70B~\cite{touvron2023llama} consumes 140GB of memory for its model weights alone (in FP16), far exceeding the capacity of a modern GPU like NVIDIA A100 or H100. This imposes a considerable challenge for LLM deployment.

To reduce inference costs in LLM deployment, low-bit quantization has become a popular approach~\cite{dettmers2022gpt3.int8,dettmers2023case,team2023gemini,young2024yi}. It reduces the precision of numerical representations of a model, thus decreasing memory footprint and computation time. 
In LLM quantization, \textit{weight quantization} is preferred over activation quantization~\cite{lin2023awq,lee2023owq}. This is because
the values of model weights are static and thus can be quantized offline. 
Weights can be quantized to 4-bit, 2-bit, and even 1-bit.
Post-training quantization (PTQ) incurs minimal accuracy loss for 4-bit weights~\cite{dettmers2023case,team2023gemini,young2024yi}. 
Recent studies and practices show that 2-bit weight quantization outperforms 4-bit in model accuracy at the same memory budget using quantization-aware training (QAT)~\cite{du2024bitdistiller,liu2025paretoq,nair2025matryoshka}.
BitNet further shows that training models with 1.58-bit (ternary) or even 1-bit (binary) weights from scratch can achieve comparable accuracy with 16-bit models~\cite{wang2023bitnet,ma2024era}.
ParetoQ\cite{liu2025paretoq}
also reports 2-bit quantization offers promising potential for memory reduction and speedup considering hardware constraints. 

Conversely, activations are generated on-the-fly with high variance, presented as dynamic outliers~\cite{dettmers2022gpt3.int8,xiao2023smoothquant,guo2023olive}. 
Due to the presence of outliers, it is challenging to quantize activations below 8 bits. Different combinations of weight and activation bit-widths have been explored across various models and scenarios~\cite{dettmers2022gpt3.int8,frantar2022gptq,guo2022ant,du2024bitdistiller,wang2023bitnet}, suggesting that no universal solution fits all scenarios. 

\subsection{LUT-based mpGEMM for Low-Bit LLM}

\begin{figure}[t]
    \centering
    \includegraphics[width=0.9\linewidth]{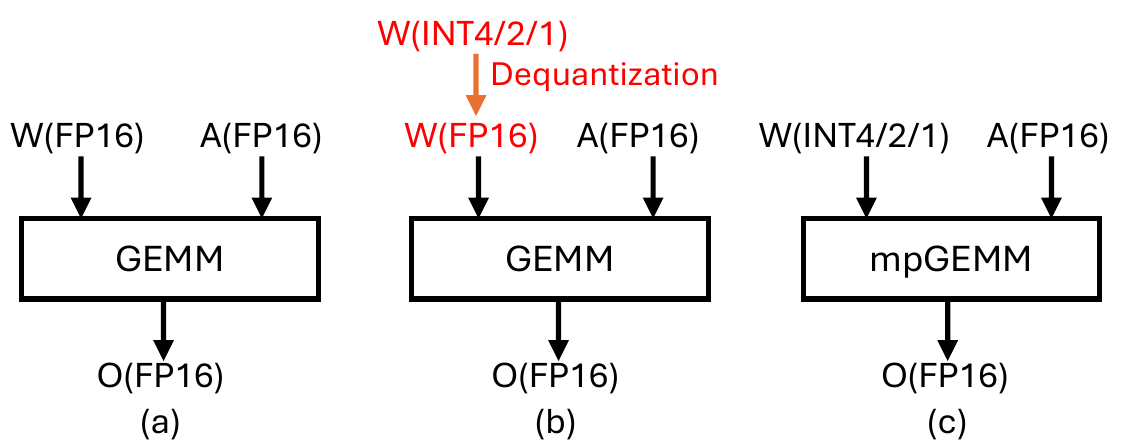}
    \vspace{-3mm}
    \caption{(a) GEMM, (b) Indirect mpGEMM with dequantization, (c) Direct mpGEMM for low-bit LLM inference.}
    \vspace{-3mm}
    \label{fig:mpGEMM}
\end{figure}

The varying bit-widths of weights and activations lead to a unique requirement of mixed-precision GEMM (mpGEMM), such as INT4/2/ 1 multiplied by FP16, illustrated in Figure~\ref{fig:mpGEMM}. Current commercial LLM inference hardware, such as GPUs and TPUs, lack native support for mpGEMM, focusing instead on conventional GEMM with uniform input formats. \textbf{Dequantization-based mpGEMM} bridges this gap by upscaling low-precision weights to match high-precision activations~\cite{cutlass,wang2024ladder}.
However, this approach introduces additional dequantization operations and forgoes the efficiency gains of low-precision computation.

\begin{figure}[t]
    \centering
    \includegraphics[width=1\linewidth]{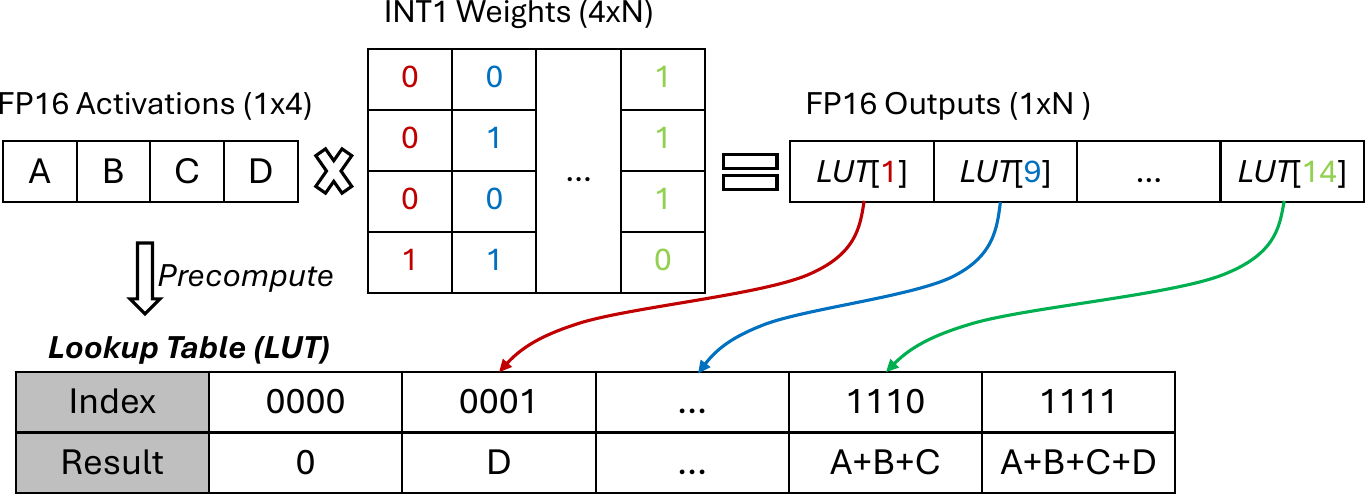}
    \vspace{-3mm}
    \caption{A naive LUT-based mpGEMM tile example of FP16 activations and INT1 weights. With the precomputed table, a table lookup can replace a dot product of 4-element vectors.}
    \vspace{-3mm}
    \label{fig:lut_gemm}
\end{figure}

\textbf{LUT-based mpGEMM} is an increasingly attractive approach for low-bit LLM inference~\cite{park2023lutgemm,jeon2020biqgemm,maleki2023look,wei2024t,unpu}. 
It precomputes dot products between high-precision activations and low-precision weights, which are then stored in lookup tables (LUT) for fast retrieval.
Instead of precomputing a massive table for all possible combinations of high-precision and low-precision values (e.g., FP16 $\times$ INT4, which would require a table of size ($2^{16} \times 2^4$)), LUT-based mpGEMM organizes computations in a tiled manner. For each small tile of the mpGEMM operation, i.e., each small group of activations, a LUT is precomputed specifically for these activation values and reused across weight columns. 
This approach minimizes table size and maintains efficiency by dynamically building LUTs for each tile during computation.
Figure~\ref{fig:lut_gemm} illustrates a basic example where a small tile consists of 1$\times$4 FP16 activations and 4$\times$N INT1 weights. With an activation vector length of 4, the lookup table size is 16. In this case, each result of the dot product of length 4 can be obtained through a simple table lookup. The table can be reused N times, which is significant given the size of the weight matrix. Larger activation vectors or higher-bit weights require proportionally larger lookup tables.

\subsection{Gaps in Current LUT-based Solutions} \label{ssec:challenges}

LUT-based mpGEMM is promising due to its advantages in eliminating dequantization and multiplication and reducing additions through simple table lookups.
However, 
existing software and hardware implementations face challenges and gaps:

\begin{figure}[t]
    \centering
    \includegraphics[width=\linewidth]{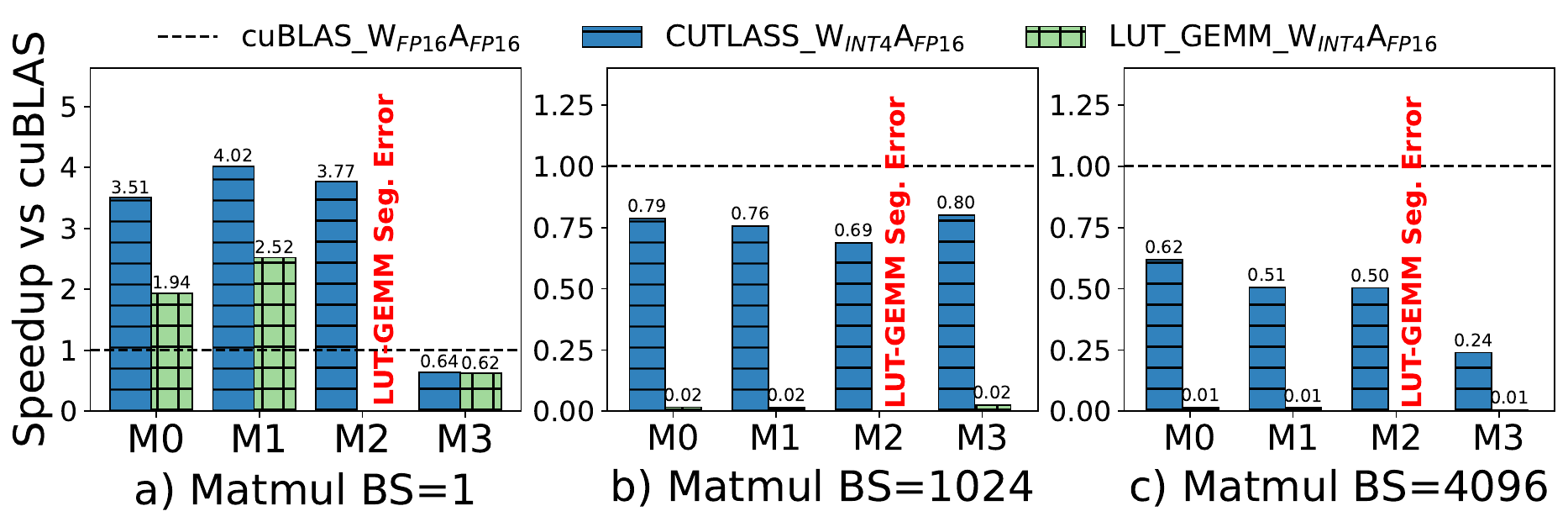}
    \vspace{-3mm}
    \caption{
    mpGEMM kernel performance with shapes M0-M3 extracted from LLAMA2-70B. $W_{INT4}A_{FP16}$ denotes INT4 weights and FP16 activations. LUT-based software kernels (LUT-GEMM) underperform dequantization-based kernels (CUTLASS) on the A100 GPU.
    }
    \vspace{-3mm}
    \label{fig:motivation}
\end{figure}

\textbf{Software LUT kernel.}
LUT-based mpGEMM software kernels often face challenges related to limited instruction support and inefficient memory access.
The limitations are two-fold: First, GPU instruction support for table lookups is limited. The most effective available instruction, \texttt{prmt} (permute), has a limited width that prevents completing a whole table lookup in a single instruction, reducing throughput. Second, table location significantly affects performance. Storing lookup tables in the register file causes extensive data duplication across threads due to the broadcast nature of LUT methods, leading to register spillage when handling large tables. Conversely, placing tables in shared memory may result in bank conflicts due to random accesses by threads within a warp, severely affecting memory bandwidth.
These issues lead to their reduced effectiveness compared to dequantization-based kernels on existing LLM inference hardware, such as GPUs.
Figure~\ref{fig:motivation} compares the performance of the LUT-based mpGEMM kernel in~\cite{park2023lutgemm} to the dequantization-based mpGEMM kernel in CUTLASS~\cite{cutlass} on A100 GPU. The results indicate that the dequantization-based kernel consistently outperforms the LUT-based kernel. 
Notably, when the batch size is large, the LUT-based kernel suffers from significant performance degradation due to table access overhead, performing several orders of magnitude worse. 
The “Seg. Error” annotation indicates a segmentation error observed in LUT-GEMM\cite{park2023lutgemm}.

\begin{figure}[t]
    \centering
    \includegraphics[width=\linewidth]{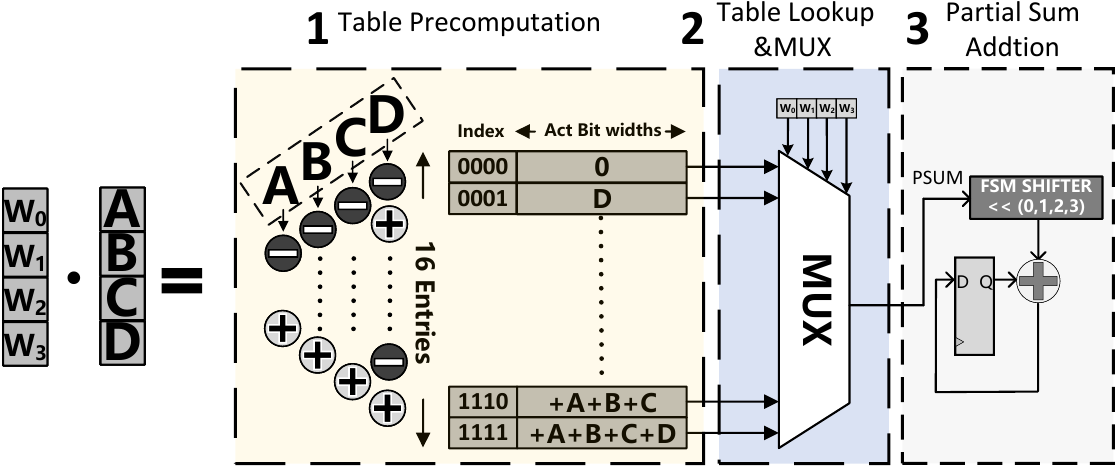}
    \vspace{-3mm}
    \caption{Conventional LUT hardware in three steps. Table precomputation and storage introduce heavy overhead.}
    \vspace{-3mm}
    \label{fig:three_step}
\end{figure}

\textbf{Hardware LUT Accelerator.}
At first glance, customized LUT hardware promises efficiency gains due to its simplicity, requiring only registers for table storage and multiplexers for lookups.
However, our study indicates that conventional LUT hardware designs fall short of delivering these gains.
Figure~\ref{fig:three_step} depicts a conventional three-step LUT-based hardware solution for mpGEMM: table precomputation, table lookup, and partial sum addition.
Numerous challenges and unexplored design aspects significantly impact the overall performance:
(1) Table precompute and storage. Precomputed tables can occupy excessive storage, incurring area and latency overhead and thus diminishing efficiency gains.
(2) Bit-width flexibility. Supporting various bit-width combinations (e.g., INT4/2/1 $\times$ FP16/FP8/INT8) may consume excessive chip area.
(3) LUT tiling shape. Suboptimal tiling increases storage costs and limits table reuse opportunities, impacting performance.
(4) Instruction and compilation. A new instruction set is required for LUT-based mpGEMM. However, the conventional compilation stack, optimized for standard GEMM hardware, may not efficiently map and schedule the new instruction set, complicating integration with existing software stacks.

\section{\oursys{} Design} \label{sec:design}

\begin{figure}[t]  
    \centering  
    \includegraphics[width=\linewidth]{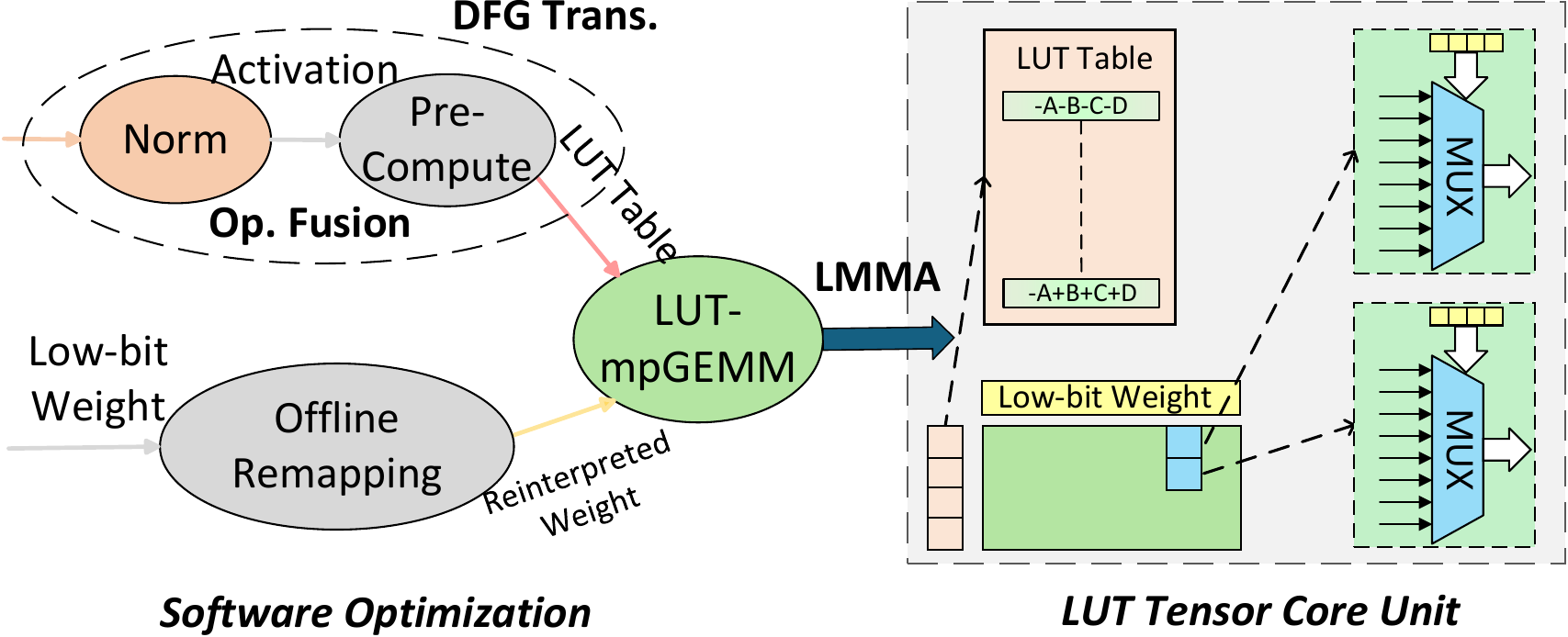}
    \caption{\oursys{} workflow.}
    \vspace{-3mm}
    \label{fig:design_overview}  
\end{figure}

We introduce \oursys{}, a software-hardware co-design 
aimed at addressing the aforementioned efficiency, flexibility, and integration challenges (\S\ref{ssec:challenges}).
Figure~\ref{fig:design_overview} provides an overview of \oursys{}.
Different from conventional hardware-based 
LUT solutions where table precompute and storage introduce significant hardware overhead, \oursys{} introduces software-based optimizations (\S\ref{ssec:design_software}) to optimize the table precompute and storage: precomputing the LUT table for the input activation tensor is performed by operator fusion, while the input weight tensor is reinterpreted to enable table storage optimizations.
On the hardware side, the LUT-based Tensor Core microarchitecture  (\S\ref{ssec:design_hardware})provides efficiency for mpGEMM processing and flexibility for different bit-width data types.
To integrate \oursys{} into existing deep learning ecosystem, \oursys{} designs the LMMA instruction set to expose the LUT-based Tensor Core for programming mpGEMMs and implements a compilation stack to schedule the end-to-end LLM execution (\S\ref{ssec:design_compilation}).

\subsection{Software-based Table Optimization} \label{ssec:design_software}

As introduced in \S\ref{sec:background}, LUT-based mpGEMM requires an additional table precomputation process and storage to store the precomputed results.
Naively, the precomputed dot products of a length $K$ activation vector on the $W\_BIT$ weight require $(2^{W\_BIT})^K$ entries for the table.
For each activation element, multiplying it with the ${W\_BIT}$ weight has $2^{W\_BIT}$ possible results, constructing the precompute table for this activation element.
Therefore, the precomputed table has $(2^{W\_BIT})^K$ entries for a length $K$ activation vector.
Figure~\ref{fig:lut_gemm} shows the lookup table with $2^4$ entries for $K=4, {W\_BIT}=1$.

A commonly-used optimization is bit-serial~\cite{judd2016stripes}, which represents a ${W\_BIT}$ integer as ${W}$ 1-bit integers and performs multiplication over 1-bit integers with bit shift. This paradigm can reuse the precompute table on 1-bit, and therefore reduces the table size to $2^K$.
Nonetheless, even this reduced table size entails significant hardware overhead. \oursys{} proposes dataflow graph (DFG) transformation and operator fusion to eliminate the table precomputation overhead, as well as weight reinterpretation and table quantization to reduce the table size.

\subsubsection{\textbf{Precomputing lookup table with DFG transformation and operator fusion}} \label{sssec:precompute_fusion}

The LUT-based mpGEMM requires precomputing the dot products between high-precision activations and a set of low-precision weights as a table for the later lookup operations.
Conventional implementations position the precompute unit adjacent to the LUT unit, performing table precomputation on-the-fly for each LUT unit. 
This approach introduces significant hardware costs due to redundancy, as multiple precompute units often perform identical operations.
Considering an example of [4096,12288]$\times$[12288,12288] GEMM in OPT-175B, a naive direct precompute unit shares a table across a LUT-based Tensor Core within an array size of N=4. 
In this setup, each table is computed repeatedly ($12288/4=3072$ times) by different LUT units throughout the process, imposing a significant computational burden.

To address this inefficiency, we first transform the DFG to split the precomputation into an independent kernel, enabling one-time precomputation that can be broadcasted to all LUT units.
This modification reduces the precomputation overhead by hundreds of times, making it manageable by existing vector units like CUDA Cores.
To amortize the additional memory traffic introduced by broadcasting, \oursys{} fuses the precompute operator with the preceding operator, leveraging its element-wise computation pattern, 
as shown in Figure~\ref{fig:design_overview} and detailed in \S\ref{sssec:compilation}. 
This fusion reduces memory access and 
brings precomputation overhead down to almost zero.
as evaluated in \S\ref{sec:precompute_analysis}.

\subsubsection{\textbf{Reinterpreting weight for table symmetrization}} \label{sssec:reinterpret_weight}

The $2^K$ table size of precomputing a length $K$ activation vector introduces a significant cost in both table storage and table accesses. To address this issue, we observed and leveraged the symmetrization property of the integer representation.

Assume that the originally quantized weights are represented as:
\vspace{-1mm}
\begin{equation}
   r_w = s_w(q_w - z_w)
   \vspace{-1mm}
\end{equation}
where $r_w$ is the real-valued weight, $s_w$ is the scale factor, $z_w$ is the bias, and $q_w$ is the $K$-bit integer representation.

Our goal is to map $q_w$ such that it is symmetric around zero while maintaining mathematical equivalence. To achieve this, both $s_w$ and $z_w$ must be adjusted. When mapping a uint $q_w$ to be symmetric about zero, the following adjustments are required:

\vspace{-2mm}
\begin{equation}
    q_w' = 2q_w - (2^K - 1), \quad s_w' = s_w / 2, \quad z_w' = 2z_w + 1 - 2^K
    \vspace{-1mm}
\end{equation}

\begin{figure}[t]  
	\centering  
    \vspace{-1mm}
    \includegraphics[width=\linewidth]{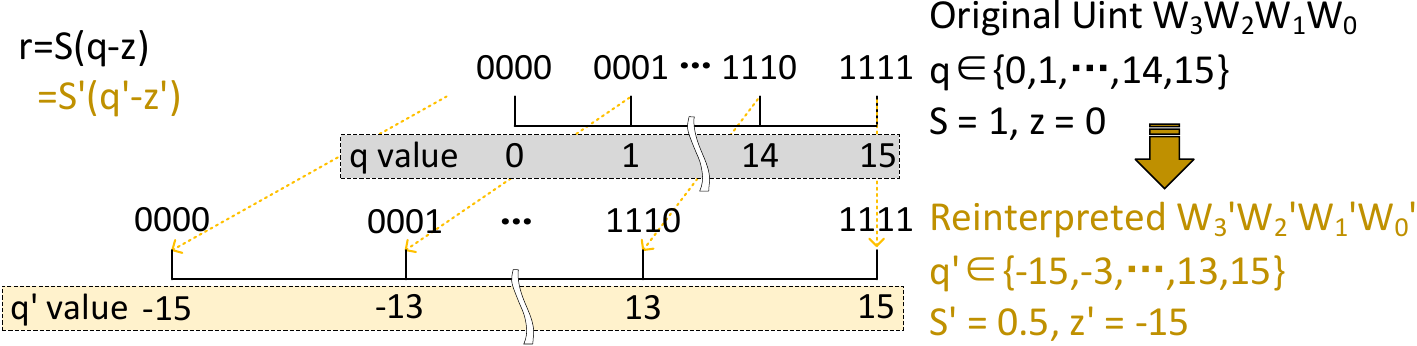}
    \vspace{-3mm}
    \caption{Reinterpreting {0,1} to {-1,1} to enable symmetry, thereby cutting the table size by half.}
    \vspace{-3mm}
    \label{fig:reinterpreted_weight}  
\end{figure}

This process is illustrated in Figure \ref{fig:reinterpreted_weight}, showing an example of transforming 4-bit unsigned integers. By calculating $s_w'$ and $z_w'$, $q_w'$ is mapped from $\{0,1,\dots, 14,15\}$ to $\{-15,-13,\dots,13,15\}$, achieving symmetry around zero.

Next, the dot product can be represented as:
\begin{equation}
    DP=\Sigma Act_is_w(q_{wi} - z_w) = \Sigma Act_is_{w}'(q_{wi}' - z_w')
\end{equation}
where $DP$ is the dot product and $Act_i$ is the activation value. Therefore, the quantization process remains the same as before, with the additional step of an offline mapping for the weight's $s_w(q_{wi} - z_w)$ to $s_w'(q_{wi}' - z_w')$.
Let us consider a dot product between the binary representation ${W_3 W_2 W_1 W_0} = {0100}$ and variables $A, B, C, D$. Initially, the binary values \{`0',`1'\} are interpreted as \{0,1\}. Assume $s_w=2$ and $z=1/2$ .The calculation proceeds as follows:
\begin{align*}
    DP &= \sum Act_i s_w \left({q}_{wi} - z_w\right) \\
       &= A \cdot 2 \cdot \left(0 - 0.5\right) + B \cdot 2 \cdot \left(1 - 0.5\right) \\ 
       & + C \cdot 2 \cdot \left(1 - 0.5\right) + D \cdot 2 \cdot \left(1 - 0.5\right) \\
       &= -A + B - C - D
\end{align*}
\vspace{-0.5mm}
After reinterpretation, the binary values \{`0',`1'\} are remapped to represent \{-1,1\}, 
with adjusted scale factor $s_w'=1$ and bias $z_w'=0$. The updated computation is:
\begin{align*}
    DP &= \sum Act_i s’_w \left({q}’_{wi} - z’_w\right) \\
       &= A \cdot 1 \cdot \left(-1 - 0\right) + B \cdot 1 \cdot \left(1 - 0\right) \\
       &+ C \cdot 1 \cdot \left(-1 - 0\right) + D \cdot 1 \cdot \left(-1 - 0\right) \\
       &= -A + B - C - D
\end{align*}
\vspace{-0.5mm}
It is clear that the two expressions remain mathematically equivalent.
As the table entries are symmetric about zero, the lookup table exhibits properties similar to odd functions. Assuming the index is a 4-bit value \( W_3 W_2 W_1 W_0 \), a naive implementation of the lookup table (LUT) requires \( 2^4 = 16 \) entries. However, it can be observed that the following property, akin to that of odd functions, holds:
\begin{equation}
    \text{LUT}[W_3 W_2 W_1 W_0] = -\text{LUT}[\sim (W_3 W_2 W_1 W_0)]
    \vspace{-1mm}
\end{equation}

Therefore, the number of entries in the LUT can be reduced to half of the original, which is \( 2^{4-1} = 8 \), and the equation becomes:
\vspace{-0.5mm}
\begin{equation}
    \text{LUT}[W_3 W_2 W_1 W_0] = 
    \begin{cases}
        -\text{LUT}[\sim (W_2 W_1 W_0)], & \text{if } W_3 = 1 \\
        \text{LUT}[W_2 W_1 W_0], & \text{if } W_3 = 0
    \end{cases}
    \label{eq:lut_formula_indexing}
    \vspace{-0.5mm}
\end{equation}
Here, $\sim$ denotes the bit-wise NOT operation. Therefore, given a length $K$ activation vector, table symmetrization can reduce the table length to \( 2^{K-1} \).
The table size not only affects the computational operations required during the precompute stage, but also the multiplexers' size. Furthermore, each entry in the table also needs to be broadcast to \( N \) PEs, typically 64 or 128, for dot product computations. Such an optimization significantly reduces the broadcasting overhead and the MUX selection overhead.
Note that \( W_3 W_2 W_1 W_0 \) in Equation~\ref{eq:lut_formula_indexing} are static weights.
The bit-level negation can be done offline to simplify the design as follows:
\vspace{-1mm}
\begin{equation}
    \text{LUT}[W_3' W_2' W_1' W_0'] = 
    \begin{cases}
        -\text{LUT}[W_2' W_1' W_0'], & \text{if } W_3' = 1 \\
        \text{LUT}[W_2' W_1' W_0'], & \text{if } W_3' = 0
    \end{cases}
    \vspace{-1mm}
    \label{eq:lut_formula_simplified}
\end{equation}
This simplification can eliminate the negation operation in circuit design, which will be introduced in \S\ref{ssec:design_hardware}.

\subsubsection{\textbf{Table quantization}}
\label{sec:table_quantization}
For high-precision activations such as FP32 or FP16, we employ table quantization techniques to convert the precomputed table elements to a lower, unified precision like INT8. This approach offers flexibility through support for multiple activation precisions and improves efficiency by reducing table size.

Compared to traditional activation quantization, table quantization allows for more dynamic, fine-grained quantization during the table precomputation stage. For example, with a group size of 4 activation elements, we perform quantization for each table that contains 8 precomputed dot products. Our empirical experiments, discussed in \S~\ref{sec:table_quantization_analysis}, demonstrate that INT8 table quantization maintains high accuracy while simplifying hardware design, thereby validating the effectiveness of our approach.

\begin{figure}[t]  
    \centering  
    \includegraphics[width=0.85\linewidth]{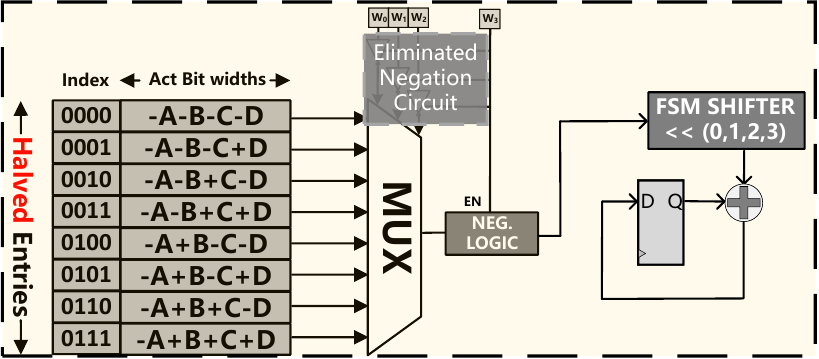}
    \vspace{-3mm}
    \caption{Optimized LUT unit with bit-serial.}
    \vspace{-3mm}
    \label{fig:lut_unit}  
\end{figure}

\subsection{LUT-based Tensor Core Microarchitecture} \label{ssec:design_hardware}

\subsubsection{\textbf{Simplified LUT unit design with bit-serial}}\label{sssec:lut_unit}
By leveraging software-based precompute fusion and weight reinterpretation, the hardware cost for customizing each individual LUT unit is reduced.
Figure~\ref{fig:lut_unit} illustrates our LUT unit design.
Compared to a straightforward design, the registers required for storing the LUT and the costs associated with table broadcasting and multiplexers are reduced by half.
As shown in Equation~\ref{eq:lut_formula_simplified}, the bit-level negation circuit can be eliminated from each LUT unit to further improve efficiency.
To support flexible bit-widths for weights, we employ a bit-serial circuit architecture~\cite{judd2016stripes,yang2021fusekna}. This design maps the weight bit-width 
to W\_BIT cycles, enabling the processing of various bit-widths in a serialized manner.

\begin{figure}[t] 
\vspace{-0.01mm}
\centering  
    \includegraphics[width=\linewidth]{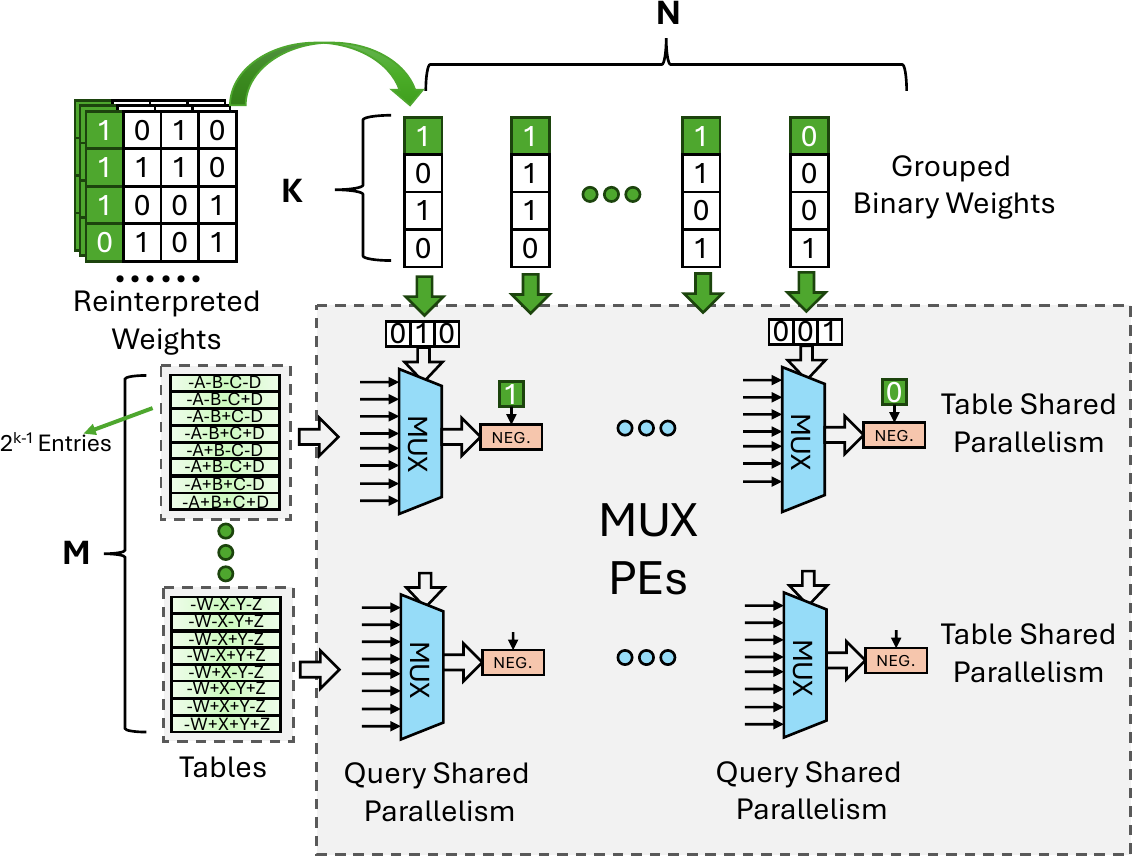}
    \vspace{-3mm}
    \caption{Elongated $MNK$ tiling of LUT-based Tensor Core. LUT-based Tensor Core requires a larger $N$ (e.g., 64/128) to maximize table reuse, along with a suitably sized $K$ (e.g., 4) for a cost-efficient table size.}
    \vspace{-5mm}
    \label{fig:lut_tiling}  
\end{figure}

\subsubsection{\textbf{Elongated LUT tiling}}\label{sssec:lut_tiling}

The selection of dimensions $M$, $N$, and $K$ is crucial for the performance of the LUT-based Tensor Core, as traditional choices for MAC-based Tensor Cores may result in suboptimal performance. As illustrated in Figure~\ref{fig:lut_tiling}, a $MNK$ Tile's LUT Array comprises $M$ tables, $N$ sets of weights, and $M*N$ MUX-based units. Each table contains \(M \times 2^{K-1}\) entries, with each entry broadcast to $N$ MUX units. Each set of \textit{Grouped Binary Weights} includes $K$ bits, which must be broadcast to $M$ MUX units to act as select signals for the MUX. The total table size is given by the equation:
\vspace{-1mm}
\begin{equation}
\text{Total Table Size} = M \times 2^{K-1} \times \text{LUT\_BIT}
    \vspace{-1mm}
\end{equation}
and the size for grouped binary weights is given by:
\vspace{-1mm}
\begin{equation}
\text{Grouped Binary Weights Size} = K \times N \times \text{W\_BIT}
    \vspace{-1mm}
\end{equation}
where \(\text{LUT\_BIT}\) is the bit width of the LUT entries, and \(\text{W\_BIT}\) is the bit width of the weights.

An LUT-based Tensor Core benefits from an elongated tiling shape. When $K$ is large, the number of table entries grows exponentially, whereas $N$ determines how many MUX units can reuse each table entry.  
An optimal configuration requires a balanced $K$, a larger $N$, and a smaller $M$, unlike conventional GPU Tensor Cores.
Additionally, the tiling shape affects I/O traffic, where a more square-like tiling configuration reduces data movement overhead. 
In \S\ref{sec:lut_array_mnk}, we explore the design space for $MNK$ tiling, confirming that elongated tiling shapes yield higher efficiency.

\subsection{Instruction and Compilation} \label{ssec:design_compilation}

To integrate \oursys{} into existing GPU architectures and ecosystems, we introduce a new instruction set and develop a compilation stack based on tile-based DNN compilers~\cite{chen2018tvm,zhu2022roller,shi2023welder}.

\begin{figure}[t]
    \centering
    \includegraphics[width=\linewidth]{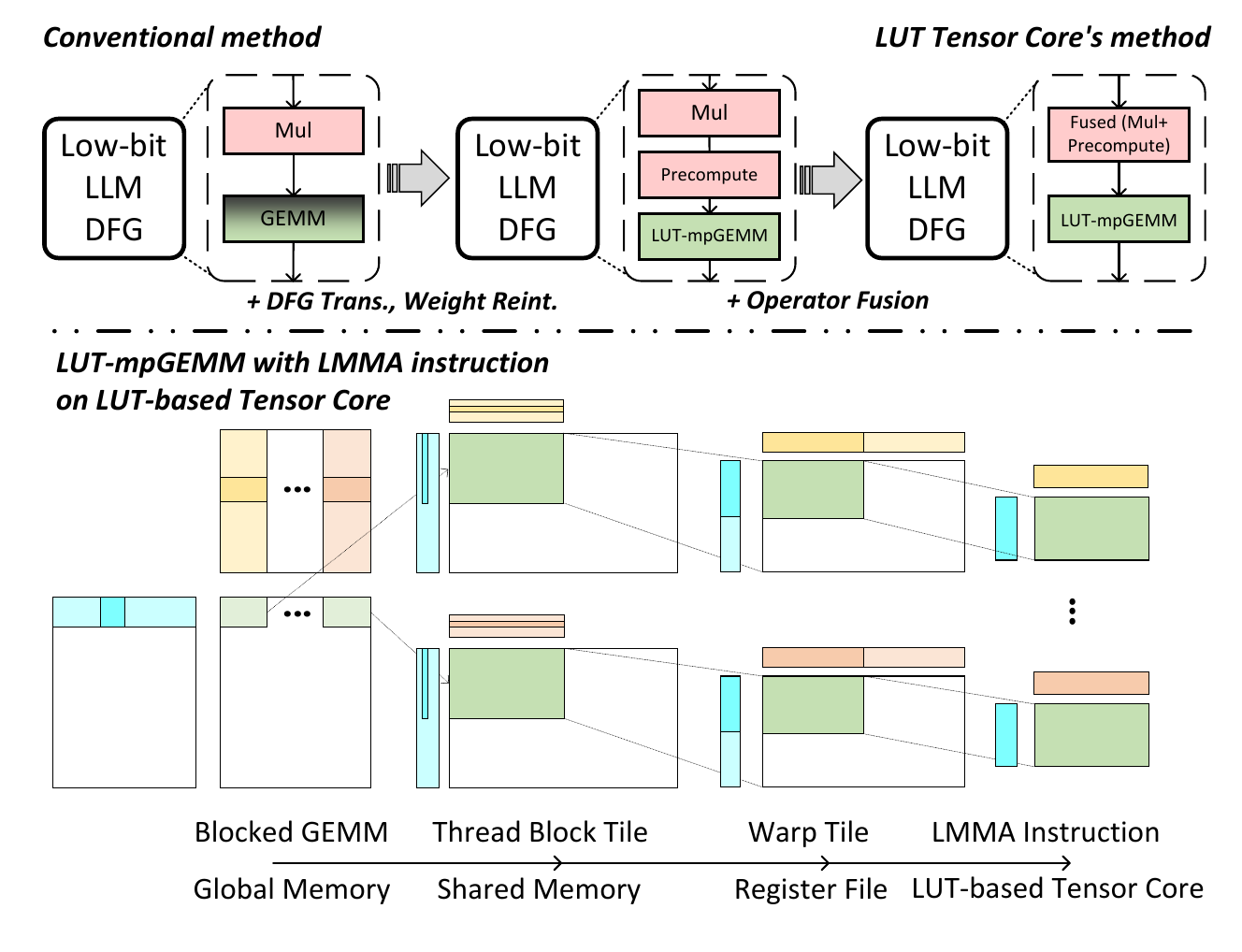}
    \vspace{-7mm}
    \caption{Compilation for LUT-mpGEMM. Overall dataflow is cutlass-like~\cite{cutlass}. Elongated tile for better data reuse.}
    \vspace{-4mm}
    \label{fig:compilation_optimization}
\end{figure}

\subsubsection{\textbf{LUT-based MMA instructions}} \label{sssec:instruction}

To enable programming with LUT-based Tensor Core, we define a set of LMMA instructions as an extension of the MMA instruction set in GPU.

\begin{center}
\textbf{lmma.\{M\}\{N\}\{K\}.\{$\boldsymbol{A_{\text{dtype}}}$\}\{$\boldsymbol{W_{\text{dtype}}}$\}\{$\boldsymbol{Accum_{\text{dtype}}}$\}\{$\boldsymbol{O_{\text{dtype}}}$\}}
\end{center}
The above formula shows the format of LMMA instructions, which resemble MMA. Specifically, the $M$, $N$, and $K$ indicate the shape of the LUT-based Tensor Core. $A_{dtype}$, $W_{dtype}$, $Accum_{dtype}$, and $O_{dtype}$ indicate the data types of the inputs, accumulation and the output, respectively.
Similar to MMA instructions, each LMMA instruction is scheduled to a warp of threads for execution. Each warp calculates the formula $O_{dtype}[M, N]$ = $A_{dtype}[M, K]$ $\times$ $W_{dtype}[N, K]$ + $Accum_{dtype}[M, N]$. 

\subsubsection{\textbf{Compilation support and optimizations}} \label{sssec:compilation}

We implemented the LUT-mpGEMM kernel generation and end-to-end LLM compilation with LUT-based Tensor Core on top of TVM~\cite{chen2018tvm}, Roller~\cite{zhu2022roller} and Welder~\cite{shi2023welder}.
Specifically, the compilation stack encompasses the following key aspects. Figure~\ref{fig:compilation_optimization} shows an example of compilation on the LLAMA model:
\begin{itemize}[leftmargin=*]
    \item \textbf{DFG Transformation.} Given the model represented in DFG, we transform the mixed-precision GEMM operator to a precompute operator and a LUT-mpGEMM operator. This transformation is implemented as a graph optimization pass in Welder~\cite{shi2023welder}.
    
    \item \textbf{Operator Fusion.}
    Operator fusion is a widely-used compiler technique to optimize the end-to-end model execution by reducing memory traffic and runtime overhead.
    We reuse Welder for operator fusion by registering the precompute and LUT-mpGEMM operators with the required tile-based representation.
    As shown in Figure~\ref{fig:compilation_optimization}, the element-wise precompute operator is fused with the previous element-wise operator.

    \item \textbf{LUT-mpGEMM Scheduling.} 
    Scheduling LUT-mpGEMM operator requires careful consideration of tiling in the memory hierarchy for optimal performance.
    Conventional GEMM tiling strategies~\cite{chen2018tvm,ansor,zhu2022roller} assume the same data type for both activations and weights. However, mpGEMM uses different data types for activation and weight, affecting memory transactions. To address this, we represent tiling by memory size rather than shape, and register LMMA instruction shapes and tiling calculations in Roller's rTile~\cite{zhu2022roller} interfaces to schedule optimal configurations.

    \item \textbf{Code Generation.}
    With the finalized scheduling plans, code generation is performed using TVM. Specifically, the LMMA instructions are registered as intrinsics in TVM, and TVM can follow the scheduling to generate the kernel code with LMMA instructions.

\end{itemize}

\section{Evaluation}

In this section, we evaluate \oursys{} to validate its efficiency in accelerating low-bit LLM inference. First, we assess the hardware efficiency gains of our design via detailed PPA benchmarking (\S\ref{ssec:eval_ppa}). Then, kernel-level experiments are conducted to illustrate the acceleration of mpGEMM (\S\ref{ssec:eval_kernel}). Next, we perform end-to-end inference evaluation on commonly-used LLMs to demonstrate the practical performance improvements (\S\ref{ssec:eval_e2e}).  
Finally, we compare \oursys{} with previous LUT-based works (\S\ref{ssec:prio_works}) and evaluate the effectiveness of our software optimizations, focusing on table precompute fusion and table quantization (\S\ref{ssec:eval_analysis}).

\begin{figure}[t]
	\centering
    \includegraphics[width=\linewidth]{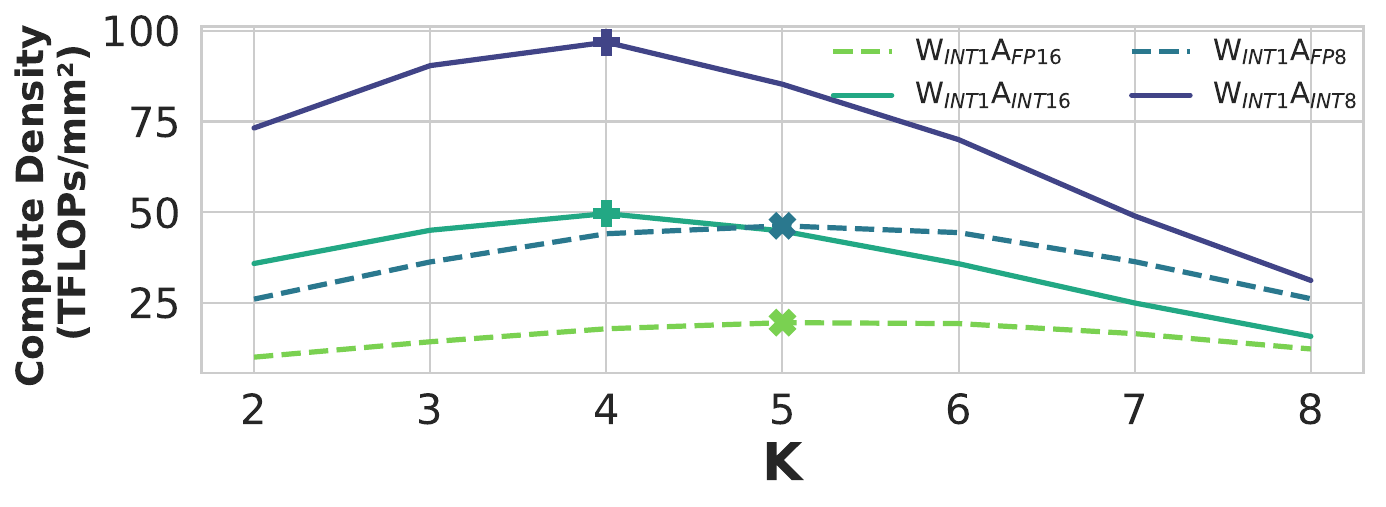}
    \vspace{-5mm}
	\caption{
    Design space exploration along the K-axis for the LUT-based dot product unit.
    K = 4 is the optimal in general.} 
    \vspace{-3mm}
	\label{fig:LUT_k_dse}
\end{figure}

\begin{figure}[t]
	\centering
    \includegraphics[width=\linewidth]{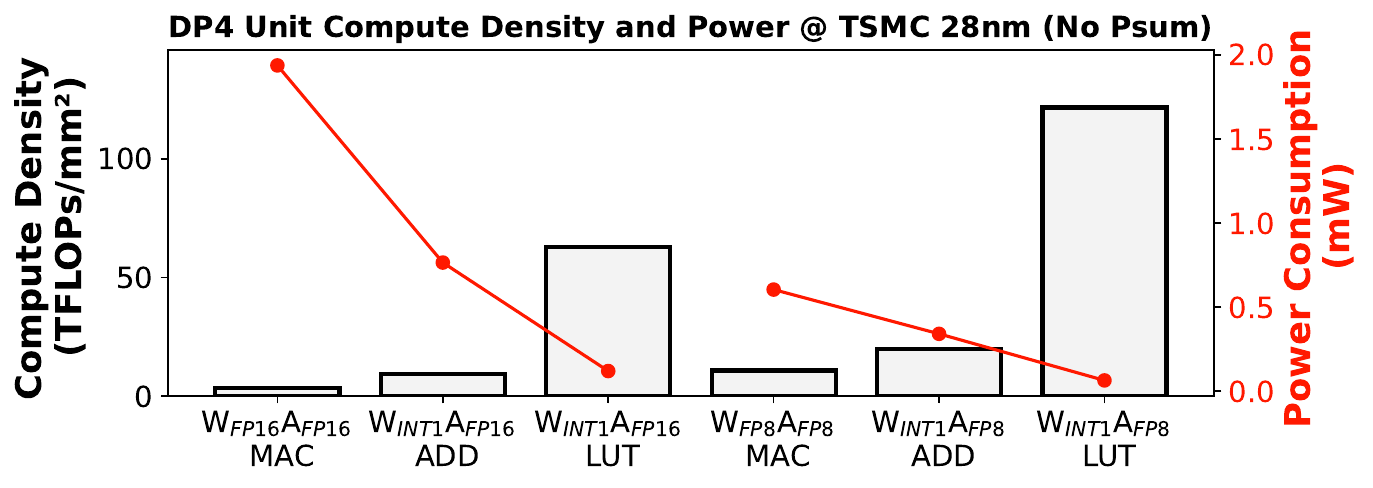}
    \vspace{-5mm}
	\caption{PPA comparison across MAC/ADD/LUT-based DP4 implementations. 
    Our LUT-based DP4 unit has compute density and power advantages.}
	\label{fig:dp4_compute_density}
\end{figure}

\begin{figure}[h]
    \centering
    \includegraphics[width=0.9\linewidth]{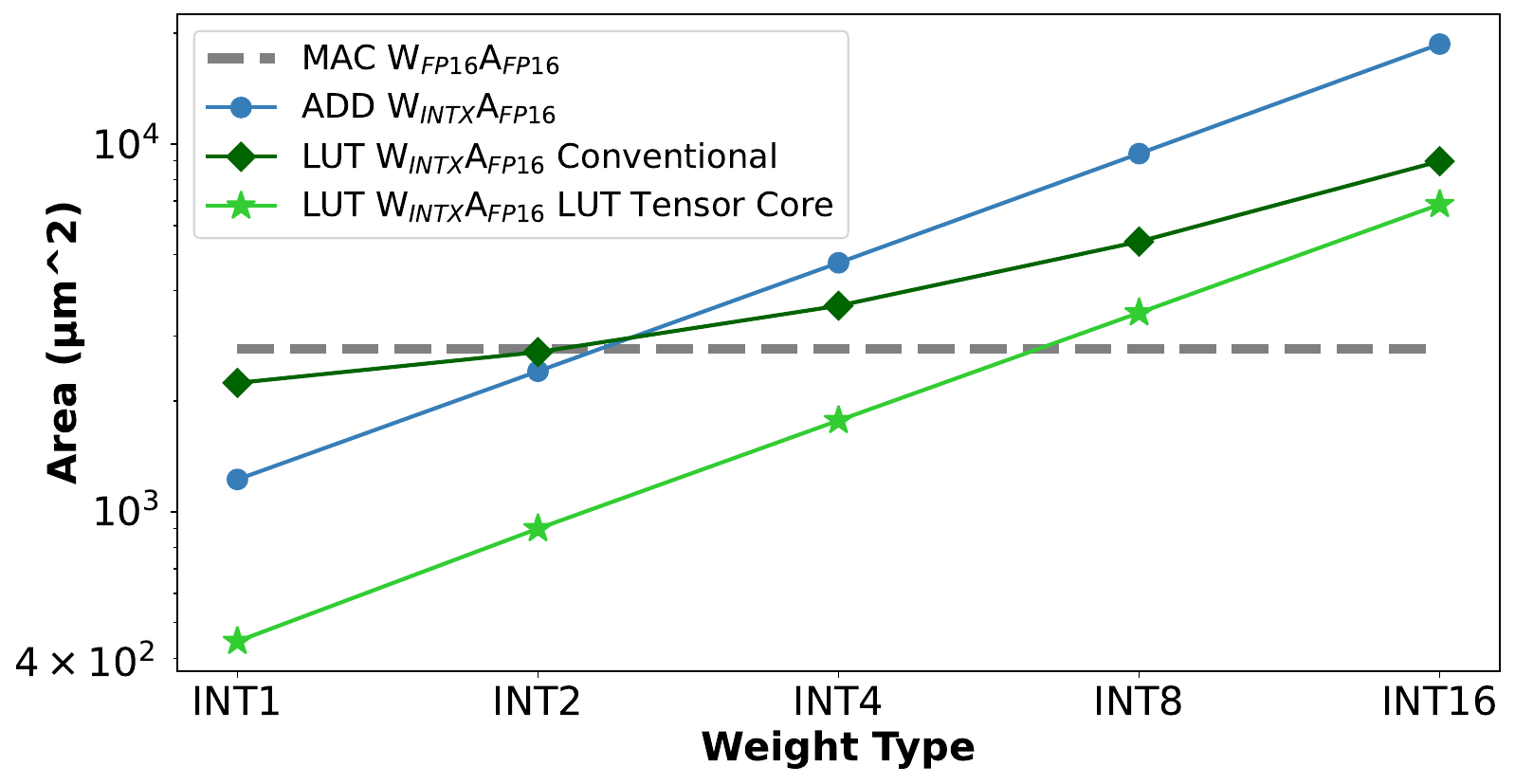}
    \vspace{-1mm}
    \caption{Area comparison of MAC, ADD, and LUT-based DP4 units across weight bit-widths in \(W_{\text{INTX}} \times A_{\text{FP16}}\). Conventional LUT implementation does not have area advantages.}
    \vspace{-2mm}
    \label{fig:dp_area_scaling_weight_bit}
\end{figure}

\subsection{Experimental Setup and Methodology} \label{ssec:eval_setup}

\subsubsection{\textbf{Hardware PPA benchmarks}}
We compare our LUT-based Tensor Core with two baselines: Multiply-Accumulate (MAC)-based Tensor Core and Addition (ADD)-based Tensor Core. MAC represents the typical design in current GPUs which needs dequantization to support mpGEMM. 
ADD adopts the bit-serial computing proposed in ~\cite{judd2016stripes} to support mpGEMM, where every bit of weights needs one addition.
We implement LUT-based Tensor Core and baselines in Verilog and use Synopsys's Design Compiler~\cite{designcompiler2018} and the TSMC 28nm process library for synthesizing circuits and generating PPA data. We apply DC's medium effort level targeting 1GHz to ensure a fair comparison across all designs. 

\subsubsection{\textbf{Kernel-level evaluation}}
For mpGEMM kernel-level evaluation, we use the NVIDIA A100 GPU as the baseline and employ Accel-Sim~\cite{khairy2020accel}, an open-source state-of-the-art simulator. 
Modifications to the configuration and trace files in Accel-Sim enable us to simulate both the original A100 and the \oursys{}-equipped A100. 

\subsubsection{\textbf{Model end-to-end evaluation and analysis}}
To extend our evaluation to real LLMs, we utilize four widely-used open-source LLMs: LLAMA-2~\cite{touvron2023llama}, OPT~\cite{zhang2022opt}, BLOOM~\cite{le2023bloom}, and BitNet~\cite{wang2023bitnet}.
Since Accel-Sim becomes infeasible for end-to-end LLM experiments due to its slow simulation speed for large trace files, we develop a tile-based simulator to support end-to-end inference evaluations, as detailed in \S\ref{ssec:eval_e2e}.

\begin{figure*}[t]
	\centering
    \includegraphics[width=\linewidth]{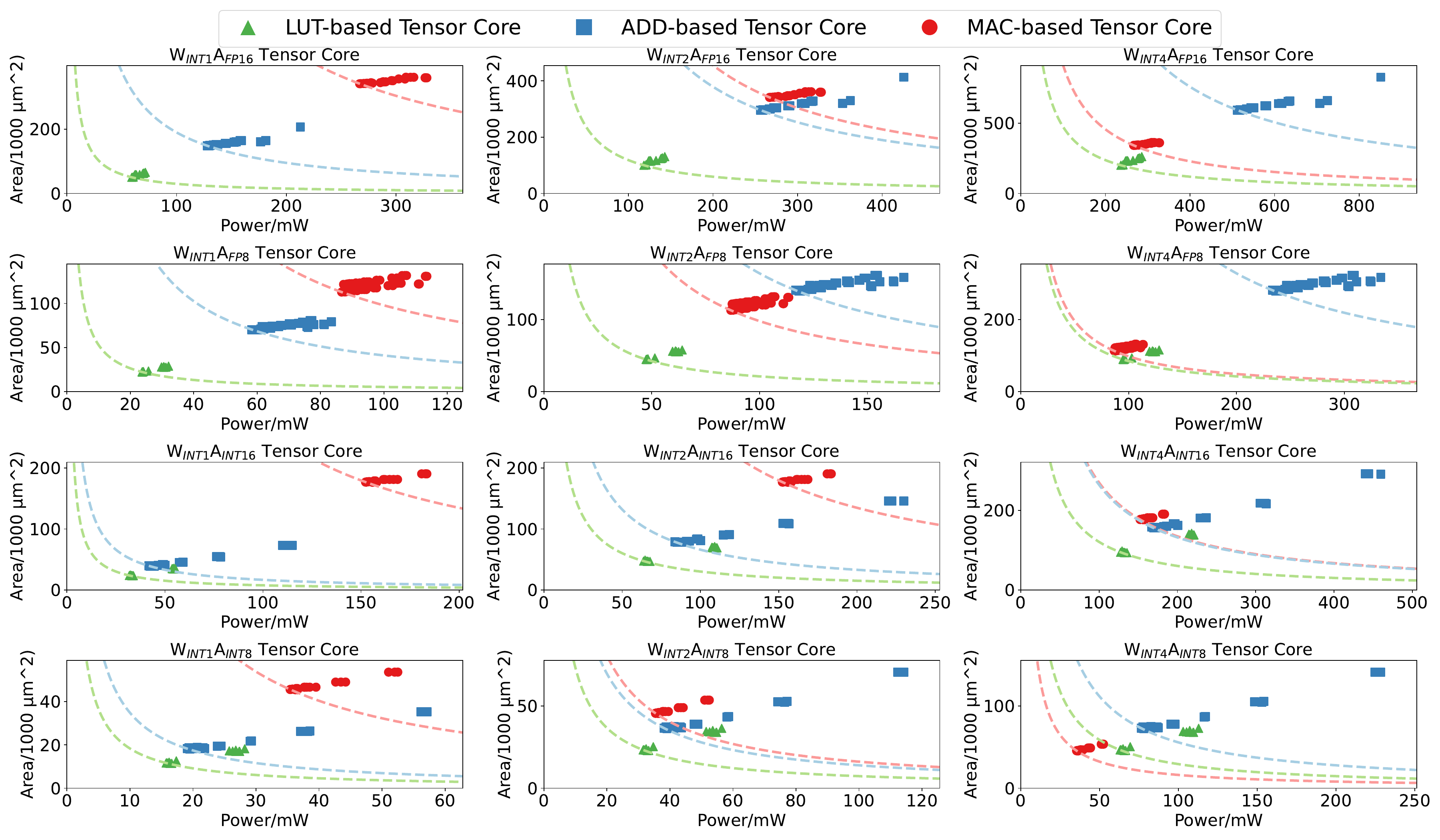}
    \vspace{-5mm}
	\caption{PPA across LUT-/ADD-/MAC-based Tensor Core implementations for mpGEMM.}
    \vspace{-2mm}
	\label{fig:a100_tensor_core_ppa}
\end{figure*}

\subsection{Hardware PPA Benchmarks} \label{ssec:eval_ppa}

\subsubsection{\textbf{Dot product unit microbenchmark}}
In this experiment, we fixed $M$ and $N$ to 1 and varied $K$ (i.e., a dot product unit of $K$-element vectors) to explore its impact on compute density. A large $K$ could lead to exponential growth in lookup table entries, whereas a smaller $K$ results in $1/K$ of the computations still being performed by adders.
As shown in Figure~\ref{fig:LUT_k_dse}, we found INT operations peak in density at $K=4$, while floating-point operations perform best at $K=5$ but also well at $K=4$. Therefore, we adopt $K=4$ for all subsequent LUT-based designs.

We conduct benchmarks on dot product implementations using MAC, ADD, and LUT-based approaches across various data formats. This includes uniform precision with MAC, such as \(W_{\text{FP16}}A_{\text{FP16}}\), and mixed precision, such as \(W_{\text{INT1}}A_{\text{FP8}}\), using both ADD and LUT approaches. As depicted in Figure~\ref{fig:dp4_compute_density}, the LUT-based approach reaches 61.55 TFLOPs/mm\(^2\) with \(W_{\text{INT1}}A_{\text{FP16}}\), surpassing the conventional MAC implementation, which only registers 3.39 TFLOPs/mm\(^2\) with \(W_{\text{FP16}}A_{\text{FP16}}\). Power efficiency shows a similar trend, with LUT-based methods achieving higher efficiency than other approaches.

Furthermore, we conduct weight-bit scaling experiments for \(W_{\text{INTX}} \times A_{\text{FP16}}\) DP4 units across MAC/ADD/LUT-based implementations. The experiments are configured with the Tensor Core's N dimension set to 4 to match the A100's configuration.
As shown in Figure~\ref{fig:dp_area_scaling_weight_bit}, the conventional LUT-based implementation does not have area advantages compared to the MAC baseline when the weight is more than 2 bits. The main area efficiency bottleneck is the table precompute and storage overhead. ADD-based implementations also only surpass the MAC baseline in the 1-bit and 2-bit cases.
Through the software-hardware co-design, \oursys{} outperforms all the baselines up to a weight bit-width of 6 and delivers better area efficiency compared to the conventional LUT implementation.

\subsubsection{\textbf{Tensor Core benchmark}}
\label{sec:lut_array_mnk}

We scale our evaluation to the Tensor Core level, incorporating a design space exploration to identify optimal $MNK$ configurations. 
To match the configuration of the A100 INT8 Tensor Core with \(M, N, K = 8, 4, 16\), we set our array size to \(M \times N \times K = 512\). 
Our experiments involve various activation data types, including \(A_{\text{FP16}}\), \(A_{\text{INT16}}\), \(A_{\text{FP8}}\), and \(A_{\text{INT8}}\), as well as multiple weight bit-widths, such as \(W_{\text{INT1}}\), \(W_{\text{INT2}}\), and \(W_{\text{INT4}}\). We compare the performance of our LUT-based approach with MAC- and ADD-based approaches.

As shown in Figure~\ref{fig:a100_tensor_core_ppa}, we sweep different $M, N, K$ configurations to explore the design space and ensure a fair comparison across all methods. The y-axis is labeled ``area'', and the x-axis is labeled ``power''.
The dashed lines represent the contours where the minimum Area$\times$Power point for each design methodology lies among all data points. Our results demonstrate that across 12 sets of experiments with different activation data formats and weight bit-widths, the LUT-based method achieves the smallest area and lowest power consumption, except in the \(W_{\text{INT8}}A_{\text{INT4}}\) case. Notably, with 1-bit weights, the LUT-based approach exhibits a 4$\times$-6$\times$ reduction in power and area compared to the MAC-based Tensor Core design.
We identify the optimal MNK configuration for the LUT-based Tensor Core as \(M2N64K4\). 
This result is due to the fact that activations are in high bits and weights are in low bits. Considering the overall bit-width, the M dimension calculates to $2 \times 16 = 32$ bits, while the N dimension computes to $64 \times 1 = 64$ bits. The overall bit configuration still approximates a square array.

\begin{figure*}[t]
    \centering
    \begin{minipage}{\linewidth}
        \centering
        \includegraphics[width=\linewidth]{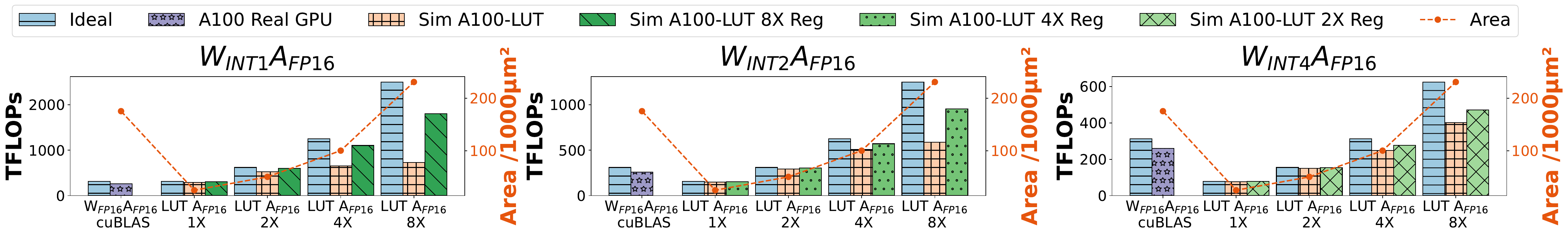}
        \label{fig:accelsim-fp16}
    \end{minipage}
    \vspace{-5mm}

    \begin{minipage}{\linewidth}
        \centering
        \includegraphics[width=\linewidth]{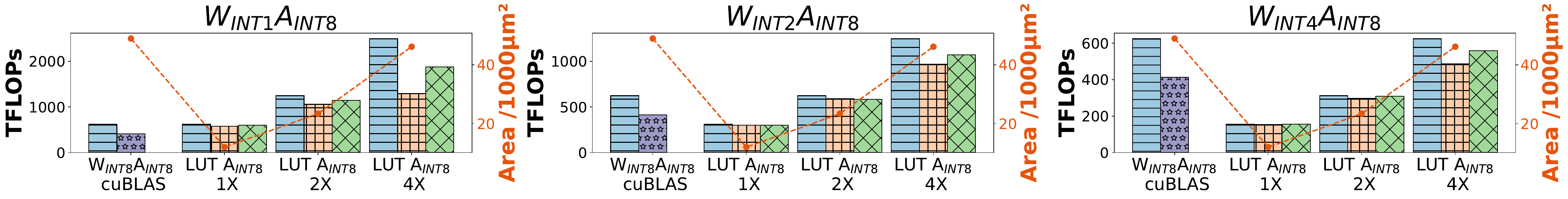}
        \label{fig:accelsim-int8}
    \end{minipage}
    \vspace{-6mm}
    \caption{Accel-Sim runtime and area across $A_{\text{FP16}}$ and $A_{\text{INT8}}$ Tensor Core designs. 
    The symbol \(\times\) denotes the Tensor Core array size relative to the 1\(\times\) baseline, where 1\(\times\) corresponds to the \(M \times N \times K = 512\) array size in the NVIDIA A100.
    }
    \label{fig:accelsim-combined}
    \vspace{-1mm}
\end{figure*}

\subsection{mpGEMM Kernel-level Evaluation} \label{ssec:eval_kernel}

We employ Accel-Sim, a SOTA GPU simulator, to validate the efficiency of \oursys{} on mpGEMM operations and its compatibility with existing GPU architectures. 
The mpGEMM shape is extracted from LLAMA2-13B, with $M=2048,N=27648,K=5120$.
The dataflow of mpGEMM is designed to be cutlass-like and output-stationary, with tiling shapes optimized for efficient data reuse. For instance, a good candidate for \(W_{\text{INT1}}A_{\text{INT8}}\) tiling sets the Thread Block tile to [128, 512, 32] and the Warp tile to [64, 256, 32].

As shown in Figure~\ref{fig:accelsim-combined}, LUT-based Tensor Core outperforms traditional MAC-based Tensor Core in mpGEMM operations. The leftmost two bars in each subplot represent A100's ideal peak performance and the measured performance using cuBLAS. The remaining bars represent LUT-based results: ideal peak performance, simulated performance, and simulated performance with an increased register capacity.
The register capacity adjustment addresses bottlenecks caused by insufficient registers, which restrict large tiling and tie performance to memory constraints. 
For example, with \(W_{\text{INT1}}A_{\text{FP16}}\), the LUT-based approach delivers slightly higher mpGEMM performance while using only 14.3\% of the area of a MAC-based Tensor Core.

\subsection{Model End-to-End Evaluation} \label{ssec:eval_e2e}

While Accel-Sim offers detailed architectural emulation, it suffers from a slowdown of approximately five million times, transforming a ten-second task on an A100 GPU into a simulation period of up to 579 days, and generating trace files over 79TB in size. 

To overcome these obstacles, we have developed an end-to-end simulator designed for rapid and accurate emulation with tile-level granularity. Our key insight is that the behavior of highly optimized, large GPU kernels with minimal stalling can be treated as accelerators, particularly in LLM scenarios. This perspective is supported by findings from NVIDIA in NVAS~\cite{villa2021need}, which suggests viewing GPU simulation philosophically as ``dynamically interacting roofline components'', rather than as a ``cycle-by-cycle progression''.
Accordingly, we adopt analytical methods from established accelerator modeling frameworks, such as Timeloop~\cite{parashar2019timeloop}, Maestro~\cite{kwon2020maestro}, and Tileflow~\cite{zheng2023tileflow}, to develop a tile-based GPU simulator. This tool facilitates a detailed and accurate evaluation of dataflow, memory bandwidth, computational resources, and operator fusion. We plan to open source this simulator in future work.

\subsubsection{\textbf{Simulator accuracy evaluation}}
\begin{figure}[t]
    \centering
    \includegraphics[width=\linewidth]{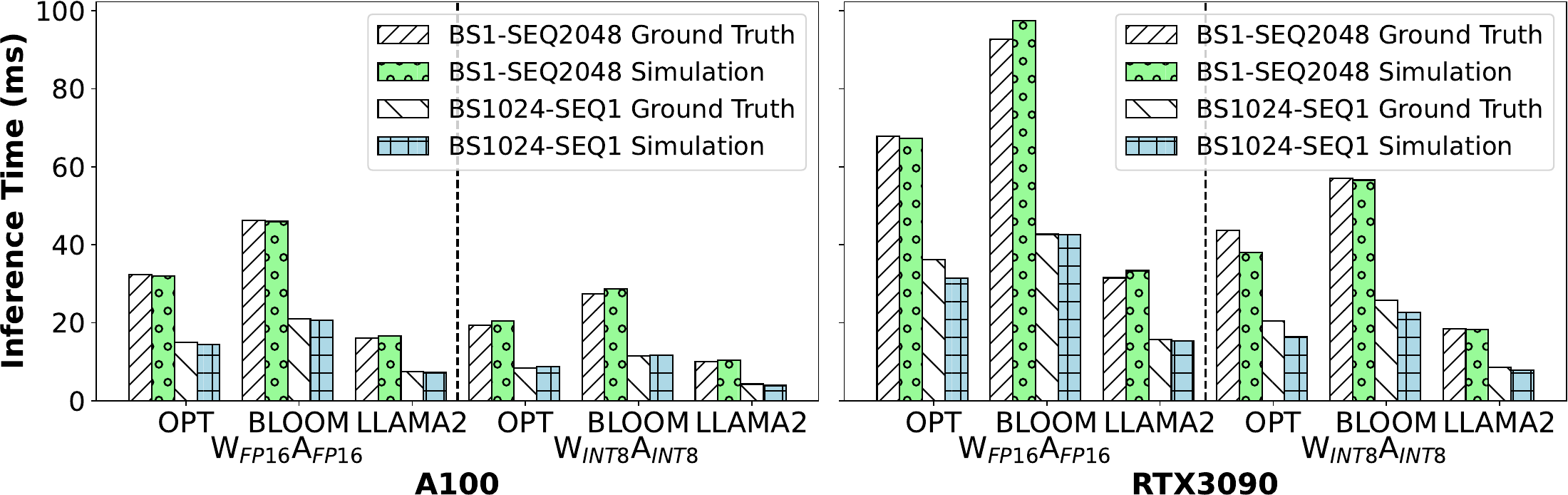}
    \vspace{-5mm}
    \caption{Evaluation of end-to-end simulator accuracy.}
    \label{fig:validation_of_e2e_modeling}
    \vspace{-1mm}
\end{figure}

In Figure~\ref{fig:validation_of_e2e_modeling}, we validate our end-to-end simulator using OPT-175B, BLOOM-176B, and LLAMA2-70B, across various configurations on a single layer on both A100 and RTX 3090 GPUs. 
Our simulator achieves a mean absolute percentage error of only 5.21\% against real GPU performance, while significantly faster than Accel-Sim in simulation speed.

\subsubsection{\textbf{End-to-end inference simulation results}}

\begin{figure}[t]
    \centering
    \includegraphics[width=\linewidth]{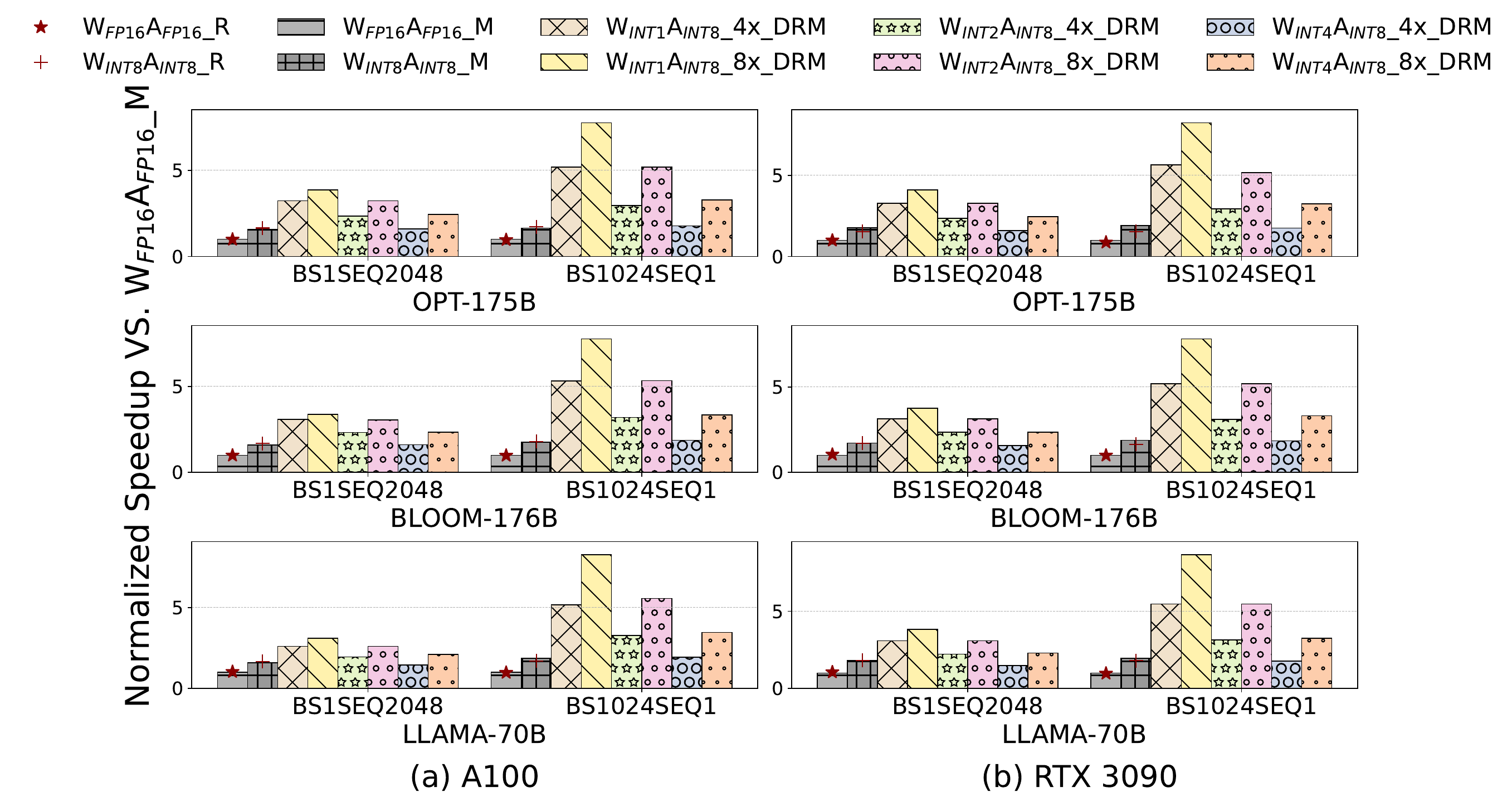}
    \vspace{-7mm}
    \caption{End-to-end simulation results on LLMs (A100 and 3090). \textit{
    R: Real GPU, M: Modeling, DRM: Double Reg Modeling.}}
    \vspace{-2mm}
    \label{fig:simulated_result_a100}
\end{figure}

Figure~\ref{fig:simulated_result_a100} presents benchmark results for the OPT, BLOOM, and LLAMA models. Our experiments demonstrate that \oursys{} achieves an end-to-end speedup of up to 8.2$\times$ while occupying less area compared to traditional $W_{\text{FP16}}A_{\text{FP16}}$ Tensor Cores. Notably, even under an 8$\times$ setting, the area of \oursys{} remains only 38.3\% that of conventional $W_{\text{FP16}}A_{\text{FP16}}$ MAC-based Tensor Cores.

\begin{table*}[t]
\centering
\phantomsection\label{CC-Q2-table} 
\caption{Overall comparison.}
\vspace{-3mm}
\label{tab:low_bit_llm_comparison}
\begin{tabular}{l|cccccccc}
\hline
\thead{\textbf{HW. Config.}} & \thead{\textbf{Model}} & \thead{\textbf{Model} \\ \textbf{Avg. Acc.}} & \thead{\textbf{BS1} \\ \textbf{SEQ2048} \\ \textbf{Latency}} & \thead{\textbf{BS1024} \\ \textbf{SEQ1} \\ \textbf{Latency}} & \thead{\textbf{Peak} \\ \textbf{Perf.}} & \thead{\textbf{TC. Area}\\ \textbf{Per SM}} & \thead{\textbf{TC. Compute} \\ \textbf{Density}} & \thead{\textbf{TC. Energy} \\ \textbf{Efficiency}} \\
\hline
A100\textsuperscript{\dag} FP16 TC. & \makecell{LLAMA 3B \\ $(W_{\text{FP16}}A_{\text{FP16}})$} & 49.7\% & 106.71ms & 41.15ms & 312 TFLOPs & 0.975mm$^2$ & 2.96 TFLOPs/mm$^2$ & 2.98 TFLOPs/W \\
A100\textsuperscript{\dag} INT8 TC & \makecell{BitNet b1.58 3B \\ $(W_{\text{INT2}}A_{\text{INT8}})$} & 49.4\% & 67.06ms & 21.70ms & 624 TOPs & 0.312mm$^2$ & 17.73 TOPs/mm$^2$ & 19.94 TOPs/W \\
A100\textsuperscript{\dag}-LUT-4X$^*$  & \makecell{BitNet b1.58 3B \\ $(W_{\text{INT2}}A_{\text{INT8}})$} & 49.4\% & 42.49ms & 11.41ms & 1248 TOPs & 0.187mm$^2$ & 61.84 TOPs/mm$^2$ & 33.32 TOPs/W \\
A100\textsuperscript{\dag}-LUT-8X$^*$  & \makecell{BitNet b1.58 3B \\ $(W_{\text{INT2}}A_{\text{INT8}})$} & 49.4\% & 38.02ms & 7.47ms & 2496 TOPs & 0.373mm$^2$ & 61.95 TOPs/mm$^2$ & 33.65 TOPs/W \\
\makecell{H100\textsuperscript{\dag} FP8 TC} & \makecell{BitNet b1.58 3B \\ $(W_{\text{FP8}}A_{\text{FP8}})$} & - & 38.20ms & 12.30ms & 1525 TFLOPs & 0.918mm$^2$ & 12.59TFLOPs/mm$^2$ & 12.24TFLOPs/W \\
\makecell{H100\textsuperscript{\dag}-LUT-4X$^*$} & \makecell{BitNet b1.58 3B \\ $(W_{\text{INT2}}A_{\text{FP8}})$} & - & 28.70ms & 9.90ms & 1525 TFLOPs & 0.488mm$^2$ & 23.69TFLOPs/mm$^2$ & 16.35TFLOPs/W \\
\makecell{H100\textsuperscript{\dag}-LUT-8X$^*$} & \makecell{BitNet b1.58 3B \\ $(W_{\text{INT2}}A_{\text{FP8}})$} & - & 23.48ms & 5.97ms & 3049 TFLOPs & 0.909mm$^2$ & 25.40TFLOPs/mm$^2$ & 17.32TFLOPs/W \\
\hline
\end{tabular}
\\  % Ensure the note starts on the next line
\vspace{1mm}
\begin{tabularx}{\textwidth}{@{}X@{}}
\textit{
Due to the lack of public data on A100/H100 Tensor Cores and their 7/4nm processes, \textsuperscript{\dag} indicates that the data are normalized to 28nm at 1.41GHz and optimized to the best of our ability for fair comparison.
-LUT$^*$ denotes \oursys{}-equipped GPU with Double Register Modeling. $\times$ means that of A100 FP16 Tensor Core array size. TC. refers to Tensor Core.  
Model accuracy for $A_{FP8}$ is not reported, as BitNet is trained from scratch in the $A_{INT8}$ format. Prior works~\cite{micikevicius2022fp8,kuzmin2022fp8,zhang2023integer} show that $A_{FP8}$ generally outperforms $A_{INT8}$ in terms of accuracy.
}
\end{tabularx}
\end{table*}

\subsubsection{\textbf{Overall comparison}}
As shown in Table~\ref{tab:low_bit_llm_comparison}, the A100 equipped with LUT + BitNet delivers up to a 5.51$\times$ acceleration in inference speed while utilizing only 38.3\% of the original Tensor Core's area. This results in an increase of up to 20.9$\times$ in compute density and an 11.2$\times$ improvement in energy efficiency, enabled by the quantized LUT table and highly optimized LUT circuit through software-hardware co-design.
Compared to the original $W_{FP8}A_{FP8}$ Tensor Core of H100, \oursys{} can achieve up to a $2.02\times$ improvement in area efficiency.

\subsection{Compared to Prior Works} \label{ssec:prio_works}

\subsubsection{\textbf{LUT-based software.}} \label{sssec:compare_lut_software}
\begin{figure}[t]
    \centering
    \includegraphics[width=\linewidth]{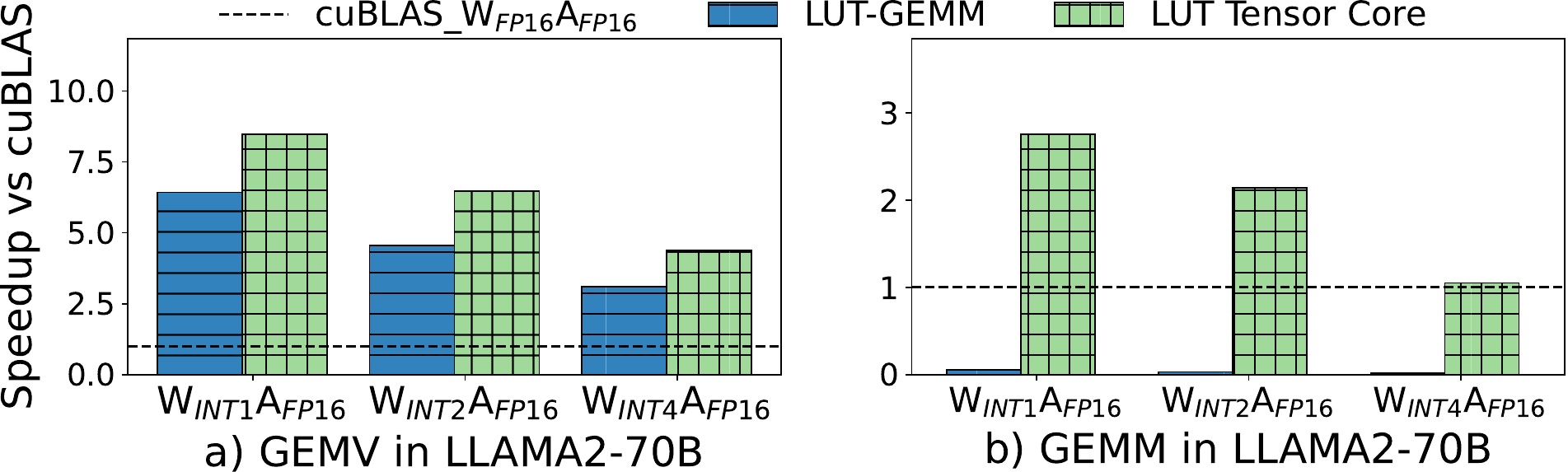}
    \vspace{-4mm}
    \caption{\oursys{} compared with LUT-based software work LUT-GEMM \cite{park2023lutgemm} for GEMM and GEMV. 
    % operations.
    }
    \vspace{-4mm}
    \label{fig:software_comparison}
\end{figure}

LUT-GEMM~\cite{park2023lutgemm} and T-MAC~\cite{wei2024t} are previous SOTA LUT-based software solutions for GPUs and CPUs, respectively. 
Since T-MAC is designed for CPUs, we use LUT-GEMM for a more relevant comparison on GPUs. \oursys{} is configured using only 57.2\% of the area of conventional FP16 Tensor Cores.
Figure~\ref{fig:software_comparison} presents the comparative speedups of \oursys{} and LUT-GEMM relative to $W_{FP16}A_{FP16}$ cuBLAS on A100.
LUT-GEMM improves performance only in GEMV cases, but is several dozen times slower in GEMM compared to cuBLAS.
Compared to the software-based LUT-GEMM, \oursys{} delivers up to 1.42$\times$ faster GEMV and 72.2$\times$ faster GEMM.

\subsubsection{\textbf{LUT-based hardware.}} \label{sssec:compare_lut_hardware}

UNPU~\cite{unpu} is the SOTA LUT-based hardware accelerator for DNN workloads. Since no public code is available, we re-implement the UNPU design based on its paper and apply optimizations to ensure a fair comparison. We conduct DSE for both UNPU and \oursys{} at the Tensor Core level. Using \(W_{\text{INT8}}A_{\text{INT2}}\) as an example under a Tensor Core configuration of \(M \times N \times K = 512\), an ablation study evaluates the impact of each optimization.
Table~\ref{tab:performance_ablation} shows that the weight reinterpretation for multi-bit weights and symmetrization enhance compute intensity and power efficiency by 30\%. 
Additional optimizations, including offline weight reinterpretation, negation circuit elimination, DFG transformation, and kernel fusion, enable \oursys{} to achieve a 1.44$\times$ improvement in these metrics compared to UNPU.

\begin{table*}[h]
\vspace{-2mm}
\centering
\caption{
\oursys{} compared with UNPU~\cite{unpu}:
\(W_{\text{INT2}}A_{\text{INT8}}\) Tensor Core case.}
\vspace{-2mm}
\label{tab:performance_ablation}
\begin{tabular}{@{}lcccc@{}}
\toprule
\textbf{Configuration} & \textbf{Area (mm\textsuperscript{2})} & \textbf{Normalized Compute Intensity} & \textbf{Power (mW)} & \textbf{Normalized Power Efficiency} \\
\midrule
UNPU (DSE Enabled)                                  & 17,271.71    & 1$\times$      & 23.39  & 1$\times$     \\
+ Weight Reinterpretation               & 13,116.60    & 1.317$\times$  & 17.98  & 1.301$\times$  \\
+ Negation Circuit Elimination                  & 12,780.05    & 1.351$\times$  & 17.37  & 1.347$\times$  \\
\makecell[tl]{+ DFG Trans. + Kernel Fusion} \\ =\oursys{} (Proposed) & 11,991.29    & 1.440$\times$  & 16.22  & 1.442$\times$  \\
\bottomrule
\end{tabular}
\end{table*}

\subsubsection{\textbf{Accelerators for quantized DNN}} \label{sssec:compare_quant_hardware}
Previous works, such as Ant~\cite{guo2022ant}, FIGNA~\cite{jang2024figna} and Mokey~\cite{Mokey}, primarily design PEs with MACs for dedicated quantized precision (e.g., int8$\times$int8 or int4$\times$fp16). While efficient for certain data types, these designs lack flexibility in adapting to different precision requirements. They either sacrifice model accuracy when converting to lower precision formats or miss efficiency opportunities when converting to higher precision formats. In contrast, we adopt a LUT-based approach that supports 1-4 bit INT weights and FP/INT 16/8 activations via different LMMA instructions, covering most low-bit LLM use cases.
Table~\ref{tab:other_accelerator} compares \oursys{} to other accelerators.

\begin{table*}[t]
\vspace{-2mm}
\centering
\caption{
\oursys{} compared with accelerators for quantized models.
}
\label{tab:other_accelerator}
\vspace{-2mm}
\resizebox{1\textwidth}{!}{

\begin{tabularx}{\textwidth}{
    >{\centering\arraybackslash}p{0.12\textwidth}  % First column
    >{\centering\arraybackslash}p{0.13\textwidth}  % Second column
    >{\centering\arraybackslash}p{0.12\textwidth}  % Third column
    >{\centering\arraybackslash}p{0.14\textwidth}  % Fourth column
    >{\centering\arraybackslash}p{0.16\textwidth}  % Fifth column
    >{\centering\arraybackslash}p{0.22\textwidth}  % Sixth column
} 
\toprule
& UNPU\cite{unpu} & Ant\cite{guo2022ant} & Mokey\cite{Mokey} & FIGNA\cite{jang2024figna} & \oursys{} \\ 
\midrule
Act. Format & INT16 & flint4 & FP16/32, INT4 & FP16/32, BF16 & FP/INT8, FP/INT16 \\

Wgt. Format & INT1$\sim$INT16 & flint4 & INT3/4 & INT4/8 & INT1$\sim$INT4 \\

Compute Engine & LUT & flint-flint MAC & Multi Counter & Pre-aligned INT MAC & LUT \\

Process & 65nm & 28nm & 65nm & 28nm & 28nm \\

PE Energy Eff. & 27TOPs/W @0.9V ($W_{\text{INT1}}A_{\text{INT16}}$) & N/A & N/A & 2.19$\times$ FP16-FP16 ($W_{\text{INT4}}A_{\text{FP16}}$) & 63.78TOPs/W @0.9V DC ($W_{\text{INT1}}A_{\text{INT8}}$) \\

Compiler Stack & \XSolidBrush & \XSolidBrush & \XSolidBrush & \XSolidBrush & \Checkmark \\
Eval. Models & VGG-16, AlexNet & ResNet, BERT & BERT, Ro/DeBERTa & BERT, BLOOM, OPT & LLAMA, BitNet, BLOOM, OPT \\
\bottomrule
\end{tabularx}
}
\end{table*}

\subsection{Software Optimization Analysis} \label{ssec:eval_analysis}

\subsubsection{\textbf{Table precompute fusion analysis}} \label{sec:precompute_analysis}

Table~\ref{tab:welder_fusion} demonstrates the impact of incorporating precomputation with the DNN compiler Welder\cite{shi2023welder}, which enhances inference performance by optimizing operator fusion. This evaluation was conducted on a single layer of the OPT-175B, BLOOM-176B, and LLAMA2-70B models in both batch prefilling and decoding configurations. Initially, precomputation on CUDA Cores led to average overhead of 16.47\% and 24.41\%. However, by treating precomputation as an independent operator within Welder’s search space, overhead is reduced to 2.62\% and 2.52\%, making it negligible in the overall execution time.

\begin{table}[t]
\vspace{-1mm}
\centering
\caption{Comparison of seperated table precompute and fused table precompute. With operator fusion, the table precompute overhead is negligible.}
\vspace{-2mm}
\label{tab:welder_fusion}
\resizebox{\columnwidth}{!}{%
\begin{tabular}{c|c|c|c|c}
\hline
\textbf{Model} & \textbf{Config} & \textbf{Welder} & \makecell{\textbf{Welder} \\ \textbf{+precompute}} & \makecell{\textbf{Welder}  \\ \textbf{+Fused precompute}} \\ \hline
OPT-175B & BS1SEQ2048 & 32.38 ms & 38.77 ms & 33.63 ms \\ 
OPT-175B & BS1024SEQ1 & 14.99 ms & 17.43 ms & 15.50 ms \\ 
BLOOM-176B & BS1SEQ4096 & 107.11 ms & 129.85 ms & 108.38 ms \\ 
BLOOM-176B & BS1024SEQ1 & 20.99 ms & 26.05 ms & 21.31 ms \\ 
LLAMA2-70B & BS1SEQ4096 & 34.68 ms & 37.60 ms & 35.65 ms \\ 
LLAMA2-70B & BS1024SEQ1 & 11.45 ms & 15.21 ms & 11.75 ms \\ \hline
\end{tabular}
}
\vspace{-2mm}
\end{table}

\subsubsection{\textbf{Table quantization analysis}} \label{sec:table_quantization_analysis}
To evaluate the impact of table quantization, we conduct comparative experiments on a LLAMA2-7B model\cite{touvron2023llama} with 2-bit quantized weights. 
The first data row represents the original $W_{FP16}A_{FP16}$ LLAMA2-7B model, and the second item corresponds to the LLAMA-3B model reported in the BitNet-b1.58 paper~\cite{ma2024era}.
The following 2-bit model is derived from BitDistiller~\cite{du2024bitdistiller}, which is an open-source QAT framework to enhance ultra low-bit LLMs. The original configuration comprised INT2 weights and FP16 activations. Building upon the open-source code of BitDistiller, we further implemented INT8 table quantization with LUT-based mpGEMM. The evaluation metrics, aligned with BitDistiller, including perplexity on the WikiText-2 dataset~\cite{merity2016pointer}, 5-shot accuracy on MMLU~\cite{hendrycks2021measuring}, and zero-shot accuracy across several tasks~\cite{zellers2019hellaswag,clark2019boolq,mihaylov2018suit,bisk2019piqa,sakaguchi2019winogrande}. The results of this empirical study are summarized in Table~\ref{tab:table_quant_analysis}. 
`N/A' in the second data row indicates that the MMLU accuracy is not reported in \cite{ma2024era}. Although the 2-bit weight quantization underperforms compared to the original $W_{FP16}A_{FP16}$ LLAMA2-7B model, it still outperforms the $W_{FP16}A_{FP16}$ LLAMA-3B model.
Notably, the INT8 table quantization does not compromise model accuracy, showing a negligible degradation in perplexity and a slight increase in task accuracy, which may be attributed to the regularizing effect of quantization.

\begin{table}[t]
\vspace{-1mm}
\centering
\caption{Table quantization analysis on LLAMA models.}
\vspace{-2mm}
\label{tab:table_quant_analysis}
\resizebox{0.5\textwidth}{!}{
\Huge % Adding \large to increase font size
\begin{tabular}{lcccccccc}
\toprule
\multirow{2}{*}{\textbf{\# Model Config.}} & 
\multirow{2}{*}{\textbf{\shortstack{WikiText2\\PPL $\downarrow$}}} & 
\multirow{2}{*}{\textbf{\shortstack{MMLU\\5s $\uparrow$}}} & 
\multicolumn{6}{c}{\textbf{Zero-shot Accuracy $\uparrow$}} \\
\cmidrule{4-9}
&&& \textbf{HS} & 
\textbf{BQ} & 
\textbf{OQ} & 
\textbf{PQ}& 
\textbf{WGe} & 
\textbf{Avg.}\\
\midrule
\phantomsection \label{rA-Q4-table-quant}
LLAMA2-7B $W_{\text{FP16}}A_{\text{FP16}}$~\cite{touvron2023llama}    & 5.47 & 45.3 & 57.1 & 77.9 & 31.4 & 78.0 & 69.1 & 62.7 \\
\addlinespace
LLAMA-3B $W_{\text{FP16}}A_{\text{FP16}}$~\cite{ma2024era}  & 10.04 & N/A  & 43.3 & 61.8 & 24.6 & 72.1 & 58.2 & 49.7 \\
\addlinespace
LLAMA2-7B $W_{\text{INT2}}A_{\text{FP16}}$~\cite{du2024bitdistiller}    & 7.68 & 30.5 & 49.2 & 70.2 & 25.8 & 73.8 & 63.1 & 56.4 \\
\addlinespace
LLAMA2-7B $W_{\text{INT2}}A_{\text{LUT\_INT8}}$~\cite{du2024bitdistiller}  & 7.69 & 30.61  & 49.2 & 70.0 & 26.2 & 73.7 & 63.5 & 56.5 \\
\bottomrule
\end{tabular}
}
\vspace{-4mm}
\end{table}

\section{Discussion and Limitations}
\label{sec:discussion}

\textbf{Low-Bit Training and Finetuning.}
Currently, \oursys{} is only applicable to 
inference acceleration for low-bit LLMs.
Recent trends show an increasing interest in low-bit training and fine-tuning for LLMs~\cite{xi2023training, dettmers2024qlora}.
While \oursys{}'s approach for mpGEMM is applicable during the forward pass of low-bit training, the complexity and stability of the training process still demand more high precision computation in the backward pass. This involves tensors and calculations such as gradients and optimizer states, which are not fully compatible with low-bit formats yet.
Further, the efficiency of training is impacted by a broad spectrum of factors such as memory efficiency and communication efficiency, beyond GEMM performance.
Consequently, optimizing the low-bit training process requires a more comprehensive strategy, possibly entailing new training algorithms that can embrace lower precision and hardware innovations to support the intricate requirements of training workflows. 
We identify these challenges as potential future directions to extend \oursys{} for training.

\textbf{Long-Context Attention and KV Cache Quantization.}
Addressing long contexts is an important frontier for LLM capabilities~\cite{peng2023yarn,ding2024longrope}.
In long-context scenarios, the attention mechanism often becomes the computational bottleneck.
Current research and practice indicate that during the prefilling stage, quantizing attention computation to FP8 does not significantly compromise model accuracy~\cite{shah2024flashattention}. However, the effects of ultra-low-bit precision on model accuracy remain largely unexplored.
During the decoding phase, several studies have shown that quantizing the KV cache to 4-bit or even 2-bit has a negligible impact on model performance~\cite{hooper2024kvquant,liu2024kivi}. Given that the Q matrix remains in high precision, the computation aligns with mpGEMM.
Exploring \oursys{} for long-context scenarios presents a promising direction for future research.

\textbf{More Data Flexibility and Non-Integer Weights.}
We believe that the LUT-based method is inherently suited for flexible precision combinations, as it replaces the main dot product operation with table lookups.
Currently, \oursys{} supports \(W_{\text{INT}}A_{\text{FP}}\) and \(W_{\text{INT}}A_{\text{INT}}\) combinations. To extend this to \(W_{\text{FP}}\), our preliminary strategy involves treating the mantissa and sign bit similarly to \(W_{\text{INT}}\), using them as table indices. The exponent bits, on the other hand, are treated as inputs to shifters. The LUT approach also accommodates non-integer weight formats. For example, in the case of ternary weights, the LUT approach can pack three ternary weights into 5 bits, whereas ADD-/MAC-based methods require 6 bits to represent the same information.

\textbf{Emerging Trends in Supporting mpGEMM.}
\label{ssec:discussion_new_hw}
Emerging GPUs such as B100~\cite{nvidia2025b100} natively support mixed-precision GEMM in Tensor Cores~\cite{cutlass,nvidia_ptx_isa}. 
Blackwell introduces narrow precision formats such as FP4, FP6, FP8, and their variants NVFP4, MXFP4, MXFP6, and MXFP8. It enables a range of mixed precision GEMM, including combinations of $A_{FP4,FP6,FP8}$ $\times$ $W_{FP4,FP6,FP8}$ and $A_{MXF4,MXF6,MXF8}$ $\times$ $W_{MXF4,MXF6,MXF8}$, while providing the same throughput as 
$W_{FP8}A_{FP8}$ Tensor Cores. 
\oursys{} supports these operations through a bit-serial approach and achieves scalable performance across different formats. 
With the emergence of native support from major vendors like NVIDIA, mpGEMM is likely to become a critical and widely-adopted computing pattern.

\begin{figure}[t]
\phantomsection\label{B-Q3} 
    \centering
    \includegraphics[width=\linewidth]{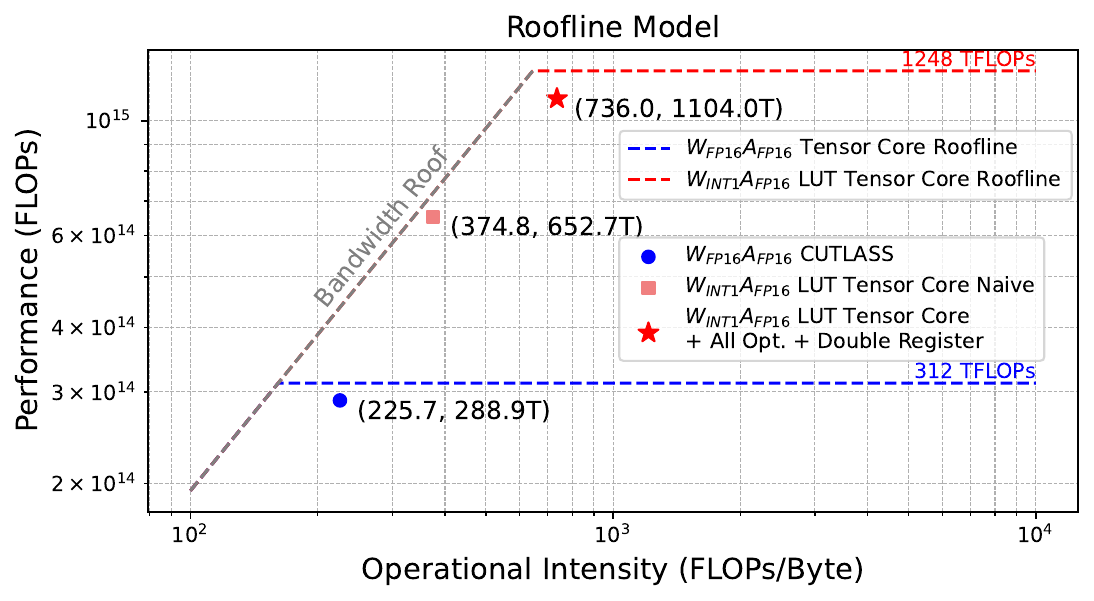}
    \vspace{-3mm}
    \caption{
    Roofline analysis of conventional $W_{FP16}A_{FP16}$ Tensor Core and $W_{INT1}A_{FP16}$ from \oursys{}.
    }
    \label{fig:roofline_analysis}
\end{figure}

\textbf{Roofline Analysis of \oursys{}.} 
Figure~\ref{fig:roofline_analysis}
presents a roofline chart for both the conventional $W_{FP16}A_{FP16}$ Tensor Core and the LUT-based $W_{INT1}A_{FP16}$ Tensor Core on an A100 memory system. The x-axis represents operational intensity based on main memory traffic. The area occupied by the $W_{INT1}A_{FP16}$ Tensor Core from \oursys{} is only 58.4\% of the area of the $W_{FP16}A_{FP16}$ Tensor Core, yet it provides 4$\times$ the theoretical FLOPs. While the original $W_{FP16}A_{FP16}$ is compute-bound, the na\"ive LUT-based implementation is memory-bound. Through the software-hardware co-optimization efforts-- reinterpreting weights to halve table size and reducing activation memory traffic, employing elongated tiling for better data reuse, and swizzling thread blocks to enhance the L2 hit rate--
\oursys{} has enhanced operational intensity and pushed the optimized point close to the "ridge point".

\vspace{-1mm}
\section{Related work}
\textbf{Low-Bit DNN Accelerators.}
As LLMs grow in size, there is an increasing need for low-bit quantization techniques to reduce model size and computational requirements. 
Hardware accelerators have been developed to efficiently support lower bit-width data types for quantized model inference.
NVIDIA's GPU architectures reflect this trend, progressively incorporating lower-precision formats. Starting with the Fermi architecture's support for FP32 and FP64, subsequent architectures have progressively included lower bit-width formats such as FP16 in Pascal, INT4 and INT8 in Turing, and BF16 in Ampere. 
In the era of LLMs, Hopper has introduced FP8~~\cite{micikevicius2022fp8} and Blackwell has advanced to FP4~\cite{rouhani2023microscaling}.
Beyond GPUs, recent studies propose customized accelerators that specifically target low-bit quantized DNNs~\cite{guo2022ant,GOBO,Bucket_MICRO23,ryu2022bitblade,Mokey,lascorz2024atalanta}. 
Although these advances demonstrate significant progress, they predominantly focus on GEMM operations where both inputs (weights and activations) share the same datatype and bit-width.
FIGNA~\cite{jang2024figna} customizes an $W_{INT4}A_{FP16}$ arithmetic unit for enhanced low-bit LLM inference. 
\oursys{} improves the efficiency of mpGEMM with LUT-based computing paradigm, and offers the flexibility to support diverse precision combinations without the need for complex hardware redesigns.

\textbf{Sparse DNN Accelerators.}
Alongside low-bit quantization, sparsity is another popular strategy to reduce model size and accelerate DNN inference. 
Sparsity leverages the inherent zero-valued elements within DNN weight matrices or activations, omitting them from computation and storage to improve efficiency.
With the advent of the NVIDIA A100 GPU, Sparse Tensor Cores were introduced, offering native support for sparsity by facilitating 2:4 structured sparsity~\cite{choquette2021nvidia}.
Beyond commercial GPUs, there has been a growing interest in customized sparse DNN accelerators.
These designs are tailored to exploit sparsity to varying degrees, often employing techniques such as pruning, zero-skipping, and sparse matrix formats to optimize both storage and computation~\cite{zhu2019sparse,wang2021dual,huang2023rm,gondimalla2023eureka,shi2024bitwave,im2024lutein,yang2021fusekna}.
Sparsity is also prevalent in low-bit LLMs. When combined with quantization, sparsity has the potential to yield even more substantial efficiency gains.
However, effectively integrating both quantization and sparsity poses significant challenges in preserving model accuracy and designing efficient microarchitectures.
Incorporating sparsity into \oursys{} represents a promising research direction, which we leave for future exploration.

\section{Conclusion}
This paper presents \oursys{}, a software-hardware co-design based on a LUT-based computing paradigm to enable efficient mixed-precision GEMM operations for low-bit LLM acceleration. 
\oursys{} enhances performance, provides broad flexibility for various precision combinations, and seamlessly integrates with existing accelerator architectures and software ecosystems.

\newpage

\bibliographystyle{ACM-Reference-Format}

% \bibliography{refs}

\end{document}